\documentclass[10pt,a4paper]{article}
\usepackage[english]{babel}

\usepackage{amsmath}
\usepackage{amsfonts}
\usepackage{amssymb}

\usepackage[colorlinks,citecolor=blue,urlcolor=blue,linkcolor=blue]{hyperref}

\usepackage[left=2cm,right=2cm,top=2cm,bottom=2cm]{geometry}

\usepackage{graphicx}

\usepackage{authblk}

\newcommand{\mP}{m_\text{P}}

\expandafter\def\csname VALIDATION_CPU_OBSERVED_MAX\endcsname{1.98407957\times10^{-12}}
\expandafter\def\csname VALIDATION_OBSERVED_MAX\endcsname{1.46670931\times10^{-9}}
\expandafter\def\csname VALIDATION_NUMERICAL_ALLOWANCE\endcsname{1.75004455\times10^{-12}}
\expandafter\def\csname VALIDATION_CPU_GPU_ALLOWANCE\endcsname{3.88020375\times10^{-7}}
\expandafter\def\csname VALIDATION_COMPLETE_BOUND\endcsname{3.89488834\times10^{-7}}
\expandafter\def\csname VALIDATION_BETA_ZERO_REGRESSION\endcsname{2.75197400\times10^{-6}}
\expandafter\def\csname VALIDATION_WRONSKIAN\endcsname{3.77687602\times10^{-7}}
\expandafter\def\csname VALIDATION_DELTA_K\endcsname{2.81645764\times10^{-8}}
\expandafter\def\csname VALIDATION_DELTA_D\endcsname{2.80264147\times10^{-8}}
\expandafter\def\csname VALIDATION_DELTA_M\endcsname{1.58572738\times10^{-11}}
\expandafter\def\csname VALIDATION_DELTA_OMEGA\endcsname{1.76090623\times10^{-10}}

\title{Robustness of Starobinsky inflation in a minimal two-field scalar-tensor completion}
\author[1,2]{Boris Latosh \thanks{latosh@theor.jinr.ru}}
\affil[1]{Bogoliubov Laboratory of Theoretical Physics, Joint Institute for Nuclear Research, 6 Joliot-Curie Street, Dubna, 141980, Russia}
\affil[2]{Dubna State University, Universitetskaya str. 19, Dubna, 141982, Russia}

\date{\today}

\begin{document}

\maketitle
\begin{abstract}
  We investigate whether the Starobinsky inflation remains robust after including a minimal scalar extension motivated by the one-loop effective action of scalar-tensor gravity. We numerically solve the four-dimensional background system derived from the reduced slow-roll action and discover that the exact Starobinsky branch is a finite-time attractor in the sampled slow-roll domain when the additional scalar is at least as heavy as the scalaron. We derive quadratic actions for tensor and scalar modes and find a healthy kinetic sector with no scalar or tensor gradient instabilities. The propagation speeds are indistinguishable from the speed of light over the tested range of the derivative-coupling coefficient. These stability tests use the complete constraint-reduced quadratic operator over $|\beta|\leq1$, whereas the quoted production curvature spectra are obtained from the reduced scalar evolution. A direct complete-spectrum comparison in $\beta$, a separate complete--reduced regression at $\beta=0$, and independent bounds on the canonically normalised coefficient differences constrain the omitted derivative-coupling correction, including the numerical and CPU--GPU allowances, to at most $\csname VALIDATION_COMPLETE_BOUND\endcsname$, below the $10^{-3}$ accuracy relevant to the physical conclusion. The curvature spectra remain extraordinarily close to the Starobinsky model across the entire parameter scan. Although the entropy-seeded mode can provide approximately one quarter of the final curvature power, removing the adiabatic--entropy mixing changes the total spectrum by only a few parts in a million, revealing nontrivial multifield dynamics behind an exceptionally robust observable prediction.
\end{abstract}


\section{Introduction}

The Starobinsky model of inflation~\cite{Starobinsky:1980te} is one of the best studied inflationary scenarios. In its simplest form, it predicts a nearly scale-invariant scalar power spectrum and a sufficiently small tensor-to-scalar ratio, both in excellent agreement with the \textit{Planck} 2018 CMB data \cite{Paoletti:2022anb,Akrami:2018odb}. At the same time, recent analyses based on the Atacama Cosmology Telescope (ACT DR6) \cite{ACT:2025fju,Ferreira:2025lrd}, especially when combined with external datasets, point towards a somewhat larger preferred value of the spectral index $n_s$ than the \textit{Planck} 2018 baseline. This does not, by itself, establish an inconsistency in the minimal Starobinsky model. However, it sharpens the motivation to test, in controlled settings, how sensitive the Starobinsky predictions are to theoretically motivated deformations. In particular, one must identify which corrections can significantly modify its predictions and which remain ineffective in a given dynamical regime.

One natural way to probe this question is to introduce quantum corrections to the classical action. Such corrections typically arise from radiative effects, matter couplings, or higher-curvature terms in the effective gravitational action (see, e.g., \cite{Sebastiani:2013eqa,Bamba:2014jia,Myrzakulov:2014hca,Myrzakulov:2016tsz,Pi:2017gih,Wang:2024vfv,Odintsov:2025eiv,Odintsov:2025jfq,Odintsov:2025zrp,Odintsov:2021wjz,Odintsov:2020ilr,Elizalde:2018now,Ben-Dayan:2014isa,Ghilencea:2018rqg,Ellis:2025bzi,Addazi:2025qra,Bianchi:2025tyl,Choi:2025qot,Wolf:2025ecy,Kim:2025dyi} for recent studies). Therefore, the recent observational situation calls for a systematic investigation of which quantum corrections can produce observable deviations from Starobinsky inflation.

The Starobinsky model occupies a special place among inflationary scenarios because it captures the ghost-free scalar part of the local curvature-squared correction to Einstein gravity that is relevant for slow-roll inflation~\cite{Ketov:2010qz,Ketov:2012jt,Buchbinder:2017lnd,Ketov:2019toi,Ketov:2025nkr}. From the EFT viewpoint, one-loop gravitational corrections generate local operators quadratic in curvature. In four dimensions, they are $R^2$ and $R_{\mu\nu}R^{\mu\nu}$ operators, plus operators that reduce to a total derivative. These two structures play qualitatively different roles: the $R^2$ term propagates an additional scalar degree of freedom (the scalaron), whereas the $R_{\mu\nu}R^{\mu\nu}$ describes a massive spin-$2$ ghost.

In this paper, we do not attempt to analyse the full quadratic-gravity EFT. We focus on the ghost-free scalar sector whose behaviour is widely studied and well understood. Concretely, we retain the $R^2$ completion and exclude the Ricci tensor squared operator, so our use of the terms ``minimal'' and ``universal'' should be understood in this restricted sense. Therefore, one can treat Starobinsky inflation as a reasonable EFT approximation. It describes the influence of one-loop effects with the $R^2$ term, while neglecting the pathological spin-2 ghost states. Although proposals exist to tame this ghost through non-standard (e.g., $PT$-symmetric) quantisation or related mechanisms \cite{Bender:2008gh,Bender:2009mq}, their status is not universally recognised, and we do not pursue them further.

We explore a way to deform the Starobinsky model within the same paradigm using operators generated by QFT at the one-loop level. We choose a model based on the following considerations. First and foremost, we focus on the scalar sector of scalar-tensor gravity. Since the standard cosmological setup admits a homogeneous and isotropic background, scalar degrees of freedom provide a natural starting point for describing the leading effects. Secondly, our goal is to isolate the gravitational radiative effects generated in the minimal scalar-tensor setup, so we deliberately switch off non-gravitational interactions of the additional scalar and keep only its minimal coupling to gravity (see, e.g., \cite{Buchbinder:2017lnd,Burgess:2003jk} for the EFT viewpoint). In this setup, the minimal extension contains one additional scalar field, which is the case studied below.

We consider the minimal scalar-tensor model developed in~\cite{Latosh:2020jyq,Arbuzov:2020pgp}. Its microscopic action and the corresponding one-loop effective action are defined canonically in Section~\ref{section_effective_action_calculation}. The effective theory retains the $R^2$ scalar sector, contains the non-minimal kinetic coupling $G^{\mu\nu}\nabla_\mu\chi\nabla_\nu\chi$ generated at the one-loop level, and includes the full one-loop potential $V(\chi)$. The auxiliary-field representation embeds into the multifield generalised $G$-inflation class of~\cite{Kobayashi:2013ina}, which makes the second-order nature of the complete field equations manifest. The parameters $m_0$ and $\beta$ are renormalised Wilson coefficients associated with the retained $R^2$ and derivative operators, while $m_\chi$ is the mass of the additional scalar.

To study slow-roll inflation in model \eqref{the_effective_action}, we proceed as follows. First, we diagonalise the model. The effective action \eqref{the_effective_action} contains two scalar degrees of freedom. The $\chi$ degree of freedom is explicitly present, while the second one -- the scalaron -- is present implicitly in full analogy with the Starobinsky model. Second, we show that the model embeds the standard Starobinsky inflation. If one sets the scalaron initial condition exactly as in the Starobinsky model and sets $\chi = \dot{\chi} =0$, then the model reproduces Starobinsky inflation.
The next question is whether nearby homogeneous trajectories are driven towards the Starobinsky branch and whether the accompanying multifield dynamics leaves observable signatures.
We study the perturbations generated during inflation to determine whether the radiative completion \eqref{the_effective_action} can produce observable deviations from the Starobinsky predictions.

The main result of this paper can be summarised as follows. The effective action \eqref{the_effective_action} contains the exact Starobinsky branch. We test its local finite-time stability by integrating the four-dimensional background equations derived from the reduced slow-roll action, together with the transverse variational equations, over a specified slow-roll domain. The sampled domain contracts uniformly for $\mu=m_\chi/m_0\geq1$, but there is a small number of finite-amplitude initial directions that do not contract over the same eight-e-fold interval for $\mu=0.90$ and $0.95$. We derive the quadratic actions for tensor and scalar modes and evaluate their kinetic, gradient, mixing, and mass sectors on every scanned background for five values of the Wilson coefficient $\beta$. No ghost or high-frequency gradient instability is found, and the scalar and tensor propagation speeds remain extremely close to unity. The reduced adiabatic--entropy system is evolved from deep inside the Hubble radius to the end of inflation. Its total curvature spectrum is practically indistinguishable from the Starobinsky model, even though the decomposition into initially adiabatic and initially entropic solutions shows nontrivial mode conversion.

The paper is organised as follows. Section~\ref{section_effective_action_calculation} introduces the microscopic scalar-tensor theory and the one-loop effective action used in the inflationary analysis. In Section~\ref{section_model_diagonalization}, we construct the auxiliary-field and diagonal scalar representations and establish the second-order nature of the complete derivative-coupling sector. Section~\ref{section_simple_inflation} studies the homogeneous slow-roll dynamics, identifies the exact Starobinsky branch, and performs the finite-time stability test using four-dimensional trajectories of the reduced background system and the transverse fundamental matrix. Section~\ref{section_perturbations} derives quadratic actions for tensor and scalar modes, presents the reduced numerical evolution of the coupled adiabatic and entropy perturbations, and tests the Bunch--Davies and entropy-transfer systematics. The results are discussed in Section~\ref{section_conclusion}.

\section{The origin of the model}\label{section_effective_action_calculation}

Multiple publications considered various quantum-corrected versions of the Starobinsky model. In most of them, the corrections are introduced phenomenologically either at the level of an effective $f(R)$ action or at the level of the scalaron potential after conformal diagonalisation. Typical examples are models with added logarithmic terms such as $R^2\ln(R/\mu^2)$, motivated by loop effects in curved space~\cite{Ben-Dayan:2014isa,Ghilencea:2018rqg}. Another approach is to extend the curvature sector beyond $R^2$ and to include higher powers of curvature~\cite{Addazi:2025qra}. Some models incorporate radiative corrections from couplings of the Starobinsky inflaton to additional matter fields via an RG-improved effective potential~\cite{Ellis:2025bzi}. Another approach is to assume that a consistent gravitational EFT also contains the Weyl-squared term and to study its impact on inflationary observables~\cite{Bianchi:2025tyl}. Related analyses argue that small corrections can substantially shift predictions of Starobinsky-like attractor models into the ACT-preferred region~\cite{Wolf:2025ecy}. All of these approaches are well motivated and can account for certain phenomenology of quantum effects, but they do not uniquely specify which operators must appear once a particular microscopic theory is fixed.

It is crucial to note that in most of the existing literature, the dominant quantum effects are assumed to come from \emph{matter} loops. In these cases, radiative corrections arise from extra fields coupled to the scalaron. Examples include Higgs-like sectors, Yukawa couplings, or other reheating-motivated interactions~\cite{Ben-Dayan:2014isa,Ghilencea:2018rqg,Ellis:2025bzi}. In this work, we take a complementary approach. We consider only corrections induced by gravity itself. More precisely, we treat general relativity as an effective field theory applicable below the Planck scale. We already computed the one-loop effective action generated by graviton fluctuations in the minimal microscopic model specified below~\cite{Latosh:2020jyq,Arbuzov:2020pgp}. This approach allows for more rigorous testing because it is more restrictive. Once the microscopic content of a model is fixed, the structure of the one-loop operators is fixed as well. It is then clear which corrections are present and which are absent. In this sense, our setup isolates the gravitational radiative effects, rather than parametrising quantum corrections through additional matter sectors or RG-improved potentials.

Below, we follow this methodology. We fix a minimal microscopic scalar-tensor theory and treat it as an effective field theory applicable below the Planck scale. We recall the computation of the one-loop effective action generated by graviton fluctuations, without introducing phenomenological corrections or contributions from additional matter degrees of freedom. This restriction makes transparent which operators arise at one loop and keeps the setup minimal, since the non-gravitational interactions of the scalar field are deliberately neglected. In the remainder of this section, we briefly outline the calculation and state the effective action obtained in~\cite{Latosh:2020jyq,Arbuzov:2020pgp} that will be used in the inflationary analysis.

The starting point of~\cite{Latosh:2020jyq,Arbuzov:2020pgp} was the microscopic action
\begin{align}\label{the_microscopic_action}
  \mathcal{A} = \int d^4 x \sqrt{-g} \left[ \cfrac{\mP^2}{2} \, R - \cfrac{1}{2} \,g^{\mu\nu} \, \nabla_\mu\chi \, \nabla_\nu\chi - \cfrac{m_\chi^2}{2}\,\chi^2 \right].
\end{align}
Here $\mP$ is the reduced Planck mass and $\chi$ is a single additional scalar field with mass $m_\chi$. We omit self-interactions such as $\chi^4$, and do not couple $\chi$ to any other matter fields. Action \eqref{the_microscopic_action} is the microscopic action, so it is used in the generating functional for quantum theory:
\begin{align}
  \mathcal{Z} = \int \mathcal{D}[g] \, \mathcal{D}[\chi] \, \exp\Big[ ~ i\, \mathcal{A}[g,\chi] ~ \Big].
\end{align}

To study quantum effects in this model, one uses the perturbative quantum gravity framework~\cite{Burgess:2003jk,Elvang:2015rqa,Vanhove:2021zel,Travaglini:2022uwo,Bambi:2023jiz,Basile:2024oms}. Gravitational degrees of freedom are small perturbations $h_{\mu\nu}$ of the spacetime metric over a flat background $\eta_{\mu\nu}$, so the spacetime metric reads:
\begin{align}
  g_{\mu\nu} = \eta_{\mu\nu} + \kappa \, h_{\mu\nu}.
\end{align}
Here $\kappa^2 = 32\pi G_\text{N} = 4\, \mP^{-2}$ is the gravitational coupling related to the Newton constant $G_\text{N}$. At the level of the generating functional, the perturbative expansion is a simple shift of the integration variables from $g_{\mu\nu}$ to $h_{\mu\nu}$, so one can change the integration variable:
\begin{align}
  \mathcal{Z} = \int\mathcal{D}[h] \, \mathcal{D}[\chi] \, \exp\Big[ i\, \mathcal{A}[\eta_{\mu\nu} + \kappa \, h_{\mu\nu} ,\chi] \Big].
\end{align}
At the one-loop level, the local gravitational EFT generates curvature-squared operators. As noted above, we restrict the analysis to the ghost-free scalar sector, so we keep the $R^2$ contribution and do not include the $R_{\mu\nu}^2$~contribution.

The one-loop effective action for \eqref{the_microscopic_action} was derived in \cite{Latosh:2020jyq,Arbuzov:2020pgp}. Perturbation theory is organised by expanding the spacetime metric in $h_{\mu\nu}$ and evaluating the one-particle-irreducible (1PI) diagrams. The effective potential $V(\chi)$ is obtained from the momentum-independent parts of the 1PI scalar vertices, with the external scalar momenta set to zero. Equivalently, this calculation determines the coefficients of the local expansion in powers of $\chi$. These coefficients are real, and their resummation gives the compact expression displayed below. Derivative operators, including the Einstein tensor kinetic coupling, vanish in the zero-momentum limit and therefore do not contribute to $V(\chi)$.

The resulting one-loop effective action takes the following compact form~\cite{Latosh:2020jyq,Arbuzov:2020pgp}
\begin{align}\label{the_effective_action}
  \Gamma = \int d^4 x \sqrt{-g}\left[ \cfrac{\mP^2}{2} \left( R + \cfrac{1}{6\, m_0^2} \, R^2 \right) - \cfrac12 \left( g^{\mu\nu} + \cfrac{4\, \beta}{\mP^2} \, G^{\mu\nu} \right) \nabla_\mu\chi \, \nabla_\nu \chi - V(\chi) \right],
\end{align}
\begin{align}
  \begin{split}\label{the_effective_potential}
    V (\chi) =&  - m_\chi^4 \,\cfrac{\ln 2}{ 64\,\pi^2} + \cfrac{m_\chi^2}{2} \left[ 1 + \cfrac{1}{8\pi^2}\, \cfrac{m_\chi^2}{\mP^2} \, (1+ \ln 4) \right] \chi^2 \\
    & + \cfrac{m_\chi^4}{128\,\pi^2} \left( 1 - 8\, \cfrac{\chi^2}{\mP^2} - \sqrt{1- 16\, \cfrac{\chi^2}{\mP^2} } \right) \ln\left[ 1 - \sqrt{ 1 - 16 \, \cfrac{\chi^2}{\mP^2} } \right] \\
    & + \cfrac{m_\chi^4}{128\,\pi^2} \left( 1 - 8\, \cfrac{\chi^2}{\mP^2} + \sqrt{1- 16\, \cfrac{\chi^2}{\mP^2} } \right) \ln\left[ 1 + \sqrt{ 1 - 16 \, \cfrac{\chi^2}{\mP^2} } \right] .
  \end{split}
\end{align}
Here $m_0$ is the scalaron mass, which is related to the renormalised coefficient of the $R^2$ operator, and $\beta$ is a dimensionless renormalised Wilson coefficient of the non-minimal kinetic coupling. The microscopic field content fixes the one-loop operator basis at the given order, whereas the finite renormalised coefficients depend on the matching and renormalisation prescription.

The resummed form \eqref{the_effective_potential} contains square roots and logarithms, so individual terms become complex when $|\chi|>\mP/4$. However, this does not imply that the whole potential is complex. The diagrammatic expansion defines a real function, and the two square-root--logarithm terms combine into the algebraically equivalent real expression
\begin{align}\label{eq:real_one_loop_potential_continuation}
  \cfrac{m_\chi^4}{128\pi^2} \Bigg[ \left(1-8\cfrac{\chi^2}{\mP^2}\right) \ln\left(16\cfrac{\chi^2}{\mP^2}\right) -2\sqrt{16\cfrac{\chi^2}{\mP^2}-1}\, \arctan\sqrt{16\cfrac{\chi^2}{\mP^2}-1} ~ \Bigg]
\end{align}
for $|\chi|>\mP/4$. Thus, the apparent restriction is a property of the chosen parametrisation, not of the potential itself. Throughout this paper, the full one-loop potential is evaluated using \eqref{the_effective_potential}, with the two square-root--logarithm terms replaced by \eqref{eq:real_one_loop_potential_continuation} for $|\chi|>\mP/4$. We do not use a truncated Taylor expansion as an approximation in either the background or perturbation calculations. The complete one-loop potential satisfies the identities, which also do not imply the usage of the Taylor expansion
\begin{align}
  V(0) &= 0, & V'(0) &= 0, & V''(0) &= m_\chi^2 .
\end{align}

The $R^2$ term describes the curvature-squared contribution retained in the effective action. The non-minimal kinetic coupling $G^{\mu\nu}\nabla_\mu\chi\nabla_\nu\chi$ is generated at the loop level in the minimal scalar-tensor setup used in this work. Despite involving higher derivatives in the action, it belongs to the Horndeski class~\cite{Horndeski:1974wa,Kobayashi:2011nu}, so the field equations remain of second order.

The important consequence of the minimal microscopic action \eqref{the_microscopic_action} is that some operators, often present in general scalar-tensor EFTs, are not generated. In particular, an $R\,\chi^2$ term does not appear in our one-loop EFT. To generate such an operator, the model must admit additional microscopic couplings, for example, a~$\chi^3$~interaction, which we have explicitly omitted (see the discussion in \cite{Latosh:2020jyq}).

\section{Model diagonalization}\label{section_model_diagonalization}

The model \eqref{the_effective_action} describes the standard gravitational degrees of freedom and two scalar degrees of freedom
\begin{align}
  \Gamma = \int d^4 x \sqrt{-g}\left[ \cfrac{\mP^2}{2} \left( R + \cfrac{1}{6\, m_0^2} \, R^2 \right) - \cfrac12 \left( g^{\mu\nu} + \cfrac{4\, \beta}{\mP^2} \, G^{\mu\nu} \right) \nabla_\mu\chi \, \nabla_\nu \chi - V(\chi) \right] .
\end{align}
To proceed, it is useful to distinguish two equivalent forms of the same theory. The first one is the auxiliary-field form, in which the $R^2$ sector is rewritten with an auxiliary scalar before the conformal transformation is performed. This form is the most convenient for identifying the Horndeski (multifield-$G$-inflation) structure and proving that the field equations are of the second order. The second one is the diagonalised form, in which the scalaron has the canonical kinetic term, and the two scalar degrees of freedom are explicit. We use the diagonalised form for the inflationary background analysis. The derivative coupling sector in the diagonalised action must always be understood as the transformed image of the complete second-order auxiliary-field theory, not as a set of independent higher-derivative operators.

Let us first introduce the auxiliary-field form of the theory. We define
\begin{align}
  f(R) = R + \cfrac{1}{6\,m_0^2} \, R^2 .
\end{align}
The higher-curvature sector can be rewritten by introducing an auxiliary scalar $\varphi$:
\begin{align}
  F(\varphi) &\overset{\text{note}}{=} f'(\varphi) = 1 + \cfrac{\varphi}{3\,m_0^2} \, , & U_\text{ aux}(\varphi) &\overset{\text{note}}{=} \cfrac{\mP^2}{2} \left[ \varphi F(\varphi)-f(\varphi) \right] = \cfrac{\mP^2}{12\,m_0^2}\,\varphi^2 .
\end{align}
Here $\varphi$ is an auxiliary scalar, and it should not be confused with the canonically normalised scalaron introduced below. The action equivalent to \eqref{the_effective_action} is then
\begin{align}\label{the_auxiliary_action}
  \Gamma_\text{ aux} = \int d^4x \sqrt{-g} \left[ \cfrac{\mP^2}{2} \, F(\varphi) R - U_\text{ aux}(\varphi) - \cfrac12 \, g^{\mu\nu}\nabla_\mu\chi\nabla_\nu\chi - \cfrac{2\beta}{\mP^2} \, G^{\mu\nu}\nabla_\mu\chi\nabla_\nu\chi - V(\chi) \right] .
\end{align}
Varying \eqref{the_auxiliary_action} with respect to $\varphi$ gives
\begin{align}
  F_{,\varphi}(\varphi)(R-\varphi)=0 .
\end{align}
Since $F_{,\varphi}=1/(3m_0^2)\neq0$, the auxiliary equation enforces $\varphi=R$.

The auxiliary-field form is useful because it makes the second-order nature of the derivative coupling manifest. Indeed, \eqref{the_auxiliary_action} embeds directly into the multifield generalised $G$-inflation class of \cite{Kobayashi:2013ina}. For several scalar fields $\phi^I$, this class is defined by the Lagrangian
\begin{align}\label{the_multifield_G_inflation_lagrangian}
  \begin{split}
    \mathcal{L}_G =& G_2(X^{IJ},\phi^K) - G_{3L}(X^{IJ},\phi^K)\,\square\phi^L \\
    & + G_4(X^{IJ},\phi^K)\,R + G_{4,\langle IJ\rangle} \left[ \square\phi^I\,\square\phi^J - \nabla_\mu\nabla_\nu\phi^I \nabla^\mu\nabla^\nu\phi^J \right] \\
    & + G_{5L}(X^{IJ},\phi^K)\, G^{\mu\nu}\nabla_\mu\nabla_\nu\phi^L \\
    & \hspace{10pt}- \cfrac{1}{6}\, G_{5I,\langle JK\rangle} \left[ \square\phi^I\,\square\phi^J\,\square\phi^K - 3\,\square\phi^{(I} \nabla_\mu\nabla_\nu\phi^J \nabla^\mu\nabla^\nu\phi^{K)} + 2\, \nabla_\mu\nabla_\nu\phi^I \nabla^\nu\nabla_\lambda\phi^J \nabla^\lambda\nabla^\mu\phi^K \right],
  \end{split}
\end{align}
where
\begin{align}
  X^{IJ} \overset{\text{note}}{=} -\cfrac12 \, g^{\mu\nu}\nabla_\mu\phi^I\nabla_\nu\phi^J .
\end{align}
The angular brackets denote the symmetrised derivative with respect to the kinetic matrix,
\begin{align}
  G_{,\langle IJ\rangle} \overset{\text{note}}{=} \cfrac12 \left( \cfrac{\partial G}{\partial X^{IJ}} + \cfrac{\partial G}{\partial X^{JI}} \right).
\end{align}
For the field equations to remain of second order, the quantities $G_{3I,\langle JK\rangle}$, $G_{4,\langle IJ\rangle,\langle KL\rangle}$, $G_{5I,\langle JK\rangle}$, $G_{5I,\langle JK\rangle,\langle LM\rangle}$ must be symmetric in all field-space indices. This condition is satisfied trivially for the embedding used below.

We now introduce
\begin{align}
  \phi^1 &\overset{\text{note}}{=} \varphi, & \phi^2 &\overset{\text{note}}{=} \chi .
\end{align}

Then \eqref{the_auxiliary_action} is reproduced by the following choice of functions:
\begin{align}
  G_2 &= X^{22} - U_\text{ aux}(\varphi) - V(\chi), & G_{31} &= 0, & G_{32} &= 0, \nonumber \\
  G_4 &= \cfrac{\mP^2}{2}F(\varphi), & G_{51} &= 0, & G_{52} &= \cfrac{2\beta}{\mP^2} \, \chi . \label{the_G_functions_for_auxiliary_action}
\end{align}
For this choice, $G_{4,\langle IJ\rangle}=0$ and $G_{5I,\langle JK\rangle}=0$, so the symmetry conditions of the multifield generalised $G$-inflation action are satisfied trivially. The term with $G_{52}$ gives
\begin{align}
  \int d^4x\sqrt{-g}\, G_{52}G^{\mu\nu}\nabla_\mu\nabla_\nu\chi &= \cfrac{2\beta}{\mP^2} \int d^4x\sqrt{-g}\, \chi\,G^{\mu\nu}\nabla_\mu\nabla_\nu\chi = -\cfrac{2\beta}{\mP^2} \int d^4x\sqrt{-g}\, G^{\mu\nu}\nabla_\mu\chi\nabla_\nu\chi \, ,
\end{align}
where we used the Bianchi identity $\nabla_\mu G^{\mu\nu}=0$ and discarded a boundary term. Thus, the Einstein tensor derivative coupling in \eqref{the_auxiliary_action} is exactly reproduced by the $G_{52}$ sector of multifield generalised $G$-inflation. Consequently, the complete field equations obtained from \eqref{the_auxiliary_action} contain at most second derivatives. This establishes the cancellation of apparent higher derivatives at the level of the full derivative coupling sector before the diagonalised form is introduced.

Having established the embedding of the auxiliary-field action into the multifield generalised $G$-inflation class of \cite{Kobayashi:2013ina}, we can now diagonalise the scalar sector. We apply the standard conformal transformation used in metric $f(R)$ gravity \cite{DeFelice:2010aj,Sotiriou:2008rp,Nojiri:2010wj,Nojiri:2017ncd}. We introduce the metric transformations
\begin{align}
  g_{\mu\nu} \to F(\varphi) \, g_{\mu\nu}.
\end{align}
This transformation is a local algebraic field redefinition: the transformed metric depends on the original fields, but not on their derivatives. Consequently, an invertible conformal transformation of this type cannot increase the differential order of the field equations. At the level of the action, however, the transformed Lagrangian may contain higher-derivative terms. This is a familiar feature of local conformal metric redefinitions in Horndeski-type theories \cite{Bettoni:2013diz,Sato:2017qau}. Such terms should not be treated as independent higher-derivative operators. They are the transformed image of the complete second-order auxiliary-field theory, and the apparent higher derivatives cancel in the full equations of motion.

The auxiliary scalar $\varphi$ is then replaced by the canonically normalised scalaron $\phi$,
\begin{align}\label{eq:model_scalar_field_redefinition}
  \phi &\overset{\text{note}}{=} \sqrt{\cfrac32}\,\mP \, \ln F(\varphi), & F(\varphi) &= \exp\left[ \sqrt{\cfrac23}\,\cfrac{\phi}{\mP} \right].
\end{align}
The higher-curvature sector then takes the canonical form
\begin{align} 
  \sqrt{-g}\, \cfrac{\mP^2}{2} \left( R + \cfrac{1}{6m_0^2} \, R^2 \right) \to \sqrt{-g} \left[ \cfrac{\mP^2}{2} \, R - \cfrac12 \, g^{\mu\nu}\nabla_\mu\phi\nabla_\nu\phi - U(\phi) \right],
\end{align}
with the standard Starobinsky potential
\begin{align}\label{eq:Starobinsky_potential}
  U(\phi) = \cfrac34 \, m_0^2 \mP^2 \left( \exp\left[ -\sqrt{\cfrac23}\,\cfrac{\phi}{\mP} \right] - 1 \right)^2 .
\end{align}
The minimally coupled kinetic term and the scalar potential transform as
\begin{align}
  \begin{split}
    -\cfrac12\sqrt{-g}\, g^{\mu\nu}\nabla_\mu\chi\nabla_\nu\chi & \to -\cfrac12\sqrt{-g}\, \exp\left[ -\sqrt{\cfrac23}\,\cfrac{\phi}{\mP} \right] g^{\mu\nu}\nabla_\mu\chi\nabla_\nu\chi, \\
    -\sqrt{-g}\,V(\chi) & \to  -\sqrt{-g}\, \exp\left[ -2\sqrt{\cfrac23}\,\cfrac{\phi}{\mP} \right] V(\chi).
  \end{split}
\end{align}

It remains to transform the Einstein tensor derivative coupling. To make the transformation transparent, we write
\begin{align}
  \sqrt{-g}\,G^{\mu\nu}\nabla_\mu\chi\nabla_\nu\chi = \cfrac12 \sqrt{-g} \left( g^{\mu\alpha}g^{\nu\beta} + g^{\mu\beta}g^{\nu\alpha} - g^{\mu\nu}g^{\alpha\beta} \right) R_{\mu\nu}\, \nabla_\alpha\chi\nabla_\beta\chi .
\end{align}
Since $\chi$ is a scalar, its first covariant derivative is independent of the connection. In four spacetime dimensions, the tensor density
\begin{align}
  \sqrt{-g} \left( g^{\mu\alpha}g^{\nu\beta} + g^{\mu\beta}g^{\nu\alpha} - g^{\mu\nu}g^{\alpha\beta} \right)
\end{align}
is conformally invariant. Therefore, the non-trivial contribution comes only from the transformation of the Ricci tensor. Therefore, the non-minimal coupling produces additional contributions only because of the Ricci tensor transformation:
\begin{align}\label{the_derivative_coupling_transform}
  \begin{split}
    \sqrt{-g}\,G^{\mu\nu}\nabla_\mu\chi\nabla_\nu\chi \to \sqrt{-g} \Bigg[ &G^{\mu\nu}\nabla_\mu\chi\nabla_\nu\chi + \sqrt{\cfrac23}\,\cfrac{1}{\mP} \left( \nabla_\mu\chi\nabla_\nu\phi\nabla^\mu\nabla^\nu\chi + \nabla_\mu\chi\nabla_\nu\chi\nabla^\mu\nabla^\nu\phi \right) \\
      &+ \cfrac{1}{6\mP^2} \left( 2\nabla_\mu\chi\nabla^\mu\phi \nabla_\nu\chi\nabla^\nu\phi + \nabla_\mu\chi\nabla^\mu\chi \nabla_\nu\phi\nabla^\nu\phi \right) \Bigg].
  \end{split}
\end{align}

Expression \eqref{the_derivative_coupling_transform} shows that the non-minimal kinetic coupling remains present after the conformal transformation. At the same time, as noted above, the transformed action also contains additional derivative interactions between the two scalar degrees of freedom, including terms with second derivatives of $\chi$ and $\phi$. Such higher-derivative terms in the transformed Lagrangian are not unexpected in Horndeski-type theories and have been discussed in the context of local conformal metric redefinitions, for example, in \cite{Bettoni:2013diz,Sato:2017qau}. Their appearance does not imply that the transformed theory has higher-order field equations. The conformal transformation used above is a local, invertible field redefinition. It contains no field derivatives and therefore cannot increase the differential order of the field equations. Consequently, the terms displayed in \eqref{the_derivative_coupling_transform} must be understood as the conformal image of the complete second-order Horndeski combination, rather than as independent higher-derivative operators. The contribution of such higher-derivative terms cancels in the equations of motion.

The diagonalised form of the effective action \eqref{the_effective_action} reads
\begin{align}\label{the_effective_action_diagonal}
  \begin{split}
    \Gamma_\text{diagonal} = \int d^4 x \sqrt{-g} \Bigg[& \cfrac{\mP^2}{2} \, R - \cfrac12\, g^{\mu\nu} \nabla_\mu \phi \, \nabla_\nu\phi- \cfrac{3}{4} \, m_0^2 \, \mP^2 \left( \exp\left[ -\sqrt{\cfrac23}\, \cfrac{\phi}{\mP} \right] - 1 \right)^2 \\
      &- \cfrac12 \, \exp\left[ - \sqrt{\cfrac23}\,\cfrac{\phi}{\mP}\right] \,  g^{\mu\nu} \nabla_\mu\chi \, \nabla_\nu \chi - \exp\left[ - 2 \sqrt{\cfrac23}\,\cfrac{\phi}{\mP}\right] V(\chi) \\
      & - \cfrac{2\, \beta}{\mP^2} \, G^{\mu\nu} \nabla_\mu\chi \, \nabla_\nu \chi - 2\, \sqrt{\cfrac23}\, \cfrac{\beta}{\mP^3} \, \left( \nabla^\mu \chi \, \nabla^\nu \phi \, \nabla_\mu \nabla_\nu \chi + \nabla^\mu \chi \, \nabla^\nu \chi \, \nabla_\mu \nabla_\nu \phi \right) \\
      & - \cfrac{\beta}{3\,\mP^4} \, \left(  2\, \nabla_\mu \chi \, \nabla^\mu \phi \, \nabla_\nu \chi \, \nabla^\nu \phi + \nabla_\mu\chi \, \nabla^\mu\chi \, \nabla_\nu\phi \, \nabla^\nu \phi \right) \Bigg].
  \end{split}
\end{align}
As discussed above, the last two lines of \eqref{the_effective_action_diagonal} contain higher-derivative terms in the transformed Lagrangian. Their role here is only to display the conformal image of the Einstein tensor derivative coupling in the auxiliary action \eqref{the_auxiliary_action} explicitly.

Let us discuss features of model \eqref{the_effective_action_diagonal} that are evident from the Lagrangian. The model extends Starobinsky inflation. Setting $\chi=0$ restores the original Starobinsky model. Below, we show that any initial conditions with $\chi(t=0)=0$ and $\dot\chi(t=0)=0$ ensure that $\chi=0$ at all times, so the model matches Starobinsky inflation exactly.

The model deviates from Starobinsky inflation if one sets non-zero initial conditions for $\chi(t=0)$ and $\dot\chi(t=0)$. The corresponding phase trajectory initially departs from the exact Starobinsky branch. The corresponding effects influence both the background solution and perturbations, thereby changing the observable power spectra.

Further, the non-minimal kinetic coupling terms are suppressed in the homogeneous slow-roll background in the vicinity of the Starobinsky branch. In a spatially flat cosmological background, these terms take the following form
\begin{align}
  \begin{split}
    - \cfrac{2\, \beta}{\mP^2} \, G^{\mu\nu} \nabla_\mu\chi \, \nabla_\nu \chi & = - 6 \,\cfrac{\beta}{\mP^2} \, H^2\, \dot\chi^2 ,\\
    - 2\, \sqrt{\cfrac23}\, \cfrac{\beta}{\mP^3} \left( \nabla^\mu \chi \, \nabla^\nu \phi \, \nabla_\mu \nabla_\nu \chi + \nabla^\mu \chi \, \nabla^\nu \chi \, \nabla_\mu \nabla_\nu \phi \right)
    &= -2 \sqrt{\cfrac23} \, \cfrac{\beta}{\mP^3}\, \dot\chi \left( \dot\phi \, \ddot\chi + \dot\chi \, \ddot\phi \right) ,\\
    - \cfrac{\beta}{3\,\mP^4} \left( 2\, \nabla_\mu \chi \, \nabla^\mu \phi \, \nabla_\nu \chi \, \nabla^\nu \phi + \nabla_\mu\chi \, \nabla^\mu\chi \, \nabla_\nu\phi \, \nabla^\nu \phi \right)
    &= -\cfrac{\beta}{\mP^4} \, \dot\phi^2\,\dot\chi^2 .
  \end{split}
\end{align}
All of them are suppressed in the slow-roll regime. In multifield inflation, one introduces several slow-roll parameters (see the discussion in \cite{Lalak:2007vi}). These describe how slowly all the fields are changing:
\begin{align}
  \epsilon_{\phi\phi} &\overset{\text{note}}{=} \cfrac{\dot\phi^2}{2\mP^2 H^2} \,, & \epsilon_{\phi\chi} &\overset{\text{note}}{=} \exp\left[ -\sqrt{\cfrac16}\,\cfrac{\phi}{\mP} \right] \cfrac{\dot\phi\,\dot\chi}{2\mP^2 H^2} \,, & \epsilon_{\chi\chi} &\overset{\text{note}}{=} \exp\left[ -\sqrt{\cfrac23}\,\cfrac{\phi}{\mP} \right] \cfrac{\dot\chi^2}{2\mP^2 H^2} \,.
\end{align}
Therefore, near the Starobinsky branch, the derivative coupling terms are suppressed by small field velocities and slow-roll accelerations. This justifies neglecting them in the background equations used below.

This statement alone, however, is not sufficient for the perturbation analysis. A term that is small on the homogeneous background can still contribute to the kinetic matrix, gradient matrix, derivative mixing, or effective masses of perturbations. For this reason, the truncation must also be checked at the level of the quadratic action for scalar and tensor modes. In Section~\ref{section_perturbations}, we obtain the perturbation action of the present model and obtain the complete tensor and scalar matrices. The resulting expressions make the no-ghost and gradient-stability conditions explicit and define the trajectory-wise test of the perturbation calculation.

Let us note that this conclusion is specific to the slow-roll scenario discussed in the present paper. It does not imply that the non-minimal derivative coupling is irrelevant in all inflationary regimes of the model~\eqref{the_effective_action}. In kinetically driven scenarios, such as $k$-inflation and $G$-inflation, derivative operators can play a leading dynamical role \cite{Armendariz-Picon:1999hyi,Garriga:1999vw, Kobayashi:2010cm,Kobayashi:2011nu}. We therefore restrict ourselves here to the slow-roll regime near the Starobinsky branch and do not analyse the possible kinetically driven solutions.

Finally, the scalar degree of freedom $\chi$ is additionally suppressed at large positive $\phi$ by the exponential prefactors in the diagonalised action
\begin{align}
  \exp\left[-\sqrt{\cfrac23}\,\cfrac{\phi}{\mP}\right] \, , & \exp\!\left[-2\sqrt{\cfrac23}\,\cfrac{\phi}{\mP}\right].
\end{align}
For Starobinsky-type initial conditions, these factors are small, so both the kinetic and potential contributions of $\chi$ to the background energy density are strongly suppressed during inflation. In this sense, $\chi$ behaves as a weakly backreacting spectator field during the initial stage of inflation. We stress, however, that this does not mean that $\chi$ is exactly decoupled from the theory. Its derivative couplings remain present in the full action and may become relevant away from the slow-roll regime or in the perturbation sector.

These features significantly simplify the analysis of inflation and cosmological perturbations in this model. We carry out this analysis in the next section.

\section{Slow-roll regime}\label{section_simple_inflation}

The presence of an attractor is essential for the predictive use of an inflationary model. The initial conditions of the very early Universe are not directly observed. Therefore, the model provides more robust predictions if it has a finite region of initial data that is dynamically driven to a single slow-roll branch. This role of slow-roll dynamics was already implicit in the original Starobinsky scenario and in the calculation of primordial perturbations in the corresponding background~\cite{Starobinsky:1980te}. It also underlies later discussions of initial conditions in inflation, stochastic inflation, and attractor-type extensions of the Starobinsky model~\cite{Mukhanov:1981xt,Starobinsky:1986fx,Carrasco:2015rva,Grain:2017dqa,Wolf:2025ecy,Ketov:2025nkr}.

In the present two-field model, the issue of attractors is more restricted. First, we need to establish whether the model has a branch that exactly matches the Starobinsky model. Second, we need to study whether this branch is locally stable against homogeneous perturbations. A global classification of the four-dimensional phase space of the reduced background system may reveal other attracting regions, but such a classification lies far beyond the scope of this paper. Consequently, we restrict the analysis to the local finite-time stability of the Starobinsky branch.

The analysis in this section uses the \textit{full one-loop potential}
\begin{align}\label{eq:attractor_total_potential}
  \mathcal{V}(\phi,\chi)  &= U(\phi)+\exp\left[-2\sqrt{\cfrac23}\,\cfrac{\phi}{\mP}\right]V(\chi)\, .
\end{align}
Here $U(\phi)$ is the Starobinsky potential \eqref{eq:Starobinsky_potential} and $V(\chi)$ is the full one-loop effective potential \eqref{the_effective_potential}.

Let us explain why it is sufficient, for the homogeneous slow-roll analysis in this section, to use the reduced two-field action. The diagonalised action \eqref{the_effective_action_diagonal} contains the canonical scalaron $\phi$, the scalar $\chi$, the potential $\mathcal{V}$, and the derivative coupling sector. The derivative coupling sector is a part of the complete theory, and one does not simply remove it from the model. However, on a spatially homogeneous cosmological background, its contributions are proportional to combinations of $\dot\chi$, $\dot\phi$, $\ddot\chi$, $\ddot\phi$, and $H^2\dot\chi^2$. In particular, as shown above, the leading homogeneous derivative coupling contributions have the form
\begin{align}\label{eq:attractor_derivative_sector_scaling}
  \begin{split}
    G^{\mu\nu}\nabla_\mu\chi\nabla_\nu\chi &\sim H^2\dot\chi^2\, , \\
    \nabla\chi\nabla\phi\nabla\nabla\chi+ \nabla\chi\nabla\chi\nabla\nabla\phi &\sim \dot\chi\left(\dot\phi\,\ddot\chi+\dot\chi\,\ddot\phi\right)\, , \\
    (\nabla\chi)^2(\nabla\phi)^2 &\sim \dot\phi^2\dot\chi^2\, .
  \end{split}
\end{align}
All these terms vanish exactly on the Starobinsky branch, where $\chi=\dot\chi=0$, and they are slow-roll suppressed in the finite region around that branch considered below. This justifies neglecting them in the homogeneous background equations used in this section. This reduction holds only at the background level, and it does not imply that the derivative coupling sector is absent from the complete theory. The influence of this sector on perturbations we discuss in Section~\ref{section_perturbations}.

Thus, for the present background analysis, we use the following reduced action
\begin{align}\label{eq:attractor_reduced_action}
  \mathcal{S}_\text{red} &= \int d^4x\sqrt{-g}\left[ \cfrac{\mP^2}{2}R -\cfrac12(\nabla\phi)^2 -\cfrac12\exp\left[-\sqrt{\cfrac23}\,\cfrac{\phi}{\mP}\right](\nabla\chi)^2 -\mathcal{V}(\phi,\chi) \right] \, .
\end{align}
We use the standard spatially flat cosmological metric
\begin{align}\label{eq:attractor_flrw_ansatz}
  ds^2 &= -dt^2+a(t)^2d\mathbf{x}^2\, , & \phi &= \phi(t)\, , & \chi &= \chi(t)\, , & H &= \cfrac{\dot a}{a}\, .
\end{align}
The corresponding field equations are
\begin{align}\label{eq:attractor_cosmic_time_equations}
  \begin{split}
    & \ddot\phi+3H\dot\phi+U'(\phi) +\cfrac{1}{\sqrt{6}\,\mP}\exp\left[-\sqrt{\cfrac23}\,\cfrac{\phi}{\mP}\right]\dot\chi^2 -2\sqrt{\cfrac23}\,\cfrac{1}{\mP}\exp\left[-2\sqrt{\cfrac23}\,\cfrac{\phi}{\mP}\right]V(\chi)=0\, , \\
    & \ddot\chi+3H\dot\chi +\exp\left[-\sqrt{\cfrac23}\,\cfrac{\phi}{\mP}\right]V'(\chi) -\sqrt{\cfrac23}\,\cfrac{\dot\phi}{\mP}\dot\chi=0\, , \\
    & 3\mP^2H^2 =\cfrac12 \, \dot\phi^2+ \cfrac12\exp\left[-\sqrt{\cfrac23}\,\cfrac{\phi}{\mP}\right]\dot\chi^2 +U(\phi)+\exp\left[-2\sqrt{\cfrac23}\,\cfrac{\phi}{\mP}\right]V(\chi)\, , \\
    & \dot H =-\cfrac{1}{2\mP^2}\left(\dot\phi^2+\exp\left[-\sqrt{\cfrac23}\,\cfrac{\phi}{\mP}\right]\dot\chi^2\right)\, .
  \end{split}
\end{align}
The Friedmann equation is a constraint. We specify the initial values of the two scalar fields and their velocities, while the positive solution of the Friedmann constraint then fixes the initial Hubble parameter.

It is convenient to use the number of e-folds as the time variable for the following analysis. The first forward crossing defines the end of inflation
\begin{align}\label{eq:attractor_end_of_inflation}
  \epsilon(N_\text{end}) &= 1\, , & \epsilon &\overset{\text{def}}{=} -\cfrac{\dot H}{H^2}\, .
\end{align}
We define the number of e-folds $N$ and the number of e-folds remaining until the end of inflation $N_\text{rem}$ as follows
\begin{align}\label{eq:attractor_efold_definitions}
  N(t) &= \ln \cfrac{a(t)}{a_\text{init}}\, , & N_\text{rem}(N) &= N_\text{end}-N\, .
\end{align}
We denote by $N_0$ the common initial surface and by $N_1$ the pivot-crossing time. The common initial scalaron state is calibrated so that the exact Starobinsky trajectory has $68$ e-folds remaining before the first crossing of $\epsilon=1$,
\begin{align}\label{eq:attractor_efold_numbers}
  N_\text{rem}(N_0) &= 68\, , & N_\text{rem}(N_1) &= 60\, .
\end{align}
We then vary the initial values of $\chi$ and $\dot\chi$ while retaining this scalaron state. These transverse displacements change the total duration only slightly: over the entire scan one finds $67.99986\leq N_\text{end}\leq68.00432$. The pivot time is located independently on every trajectory by $N_\text{rem}=60$. Consequently, the relaxation interval equals eight e-folds on the exact branch and differs from it by no more than $4.4\times10^{-3}$ e-folds on the neighbouring trajectories. This interval is used only to test homogeneous relaxation. In a two-field system, the curvature perturbation need not freeze at Hubble exit because entropy perturbations can source it on super-Hubble scales~\cite{Lalak:2007vi,Langlois:2008mn}. For this reason, Section~\ref{section_perturbations} evolves every Fourier mode to the end of inflation.

To proceed, we introduce dimensionless variables
\begin{align}\label{eq:attractor_dimensionless_variables}
  x &= \cfrac{\phi}{\mP}\, , & z &= \cfrac{\chi}{\mP}\, , & P_x &= \cfrac{dx}{dN}\, , & P_z &= \cfrac{dz}{dN}\, , & y &= \exp[-\alpha x] \, , & \alpha &= \sqrt{\cfrac23}\, .
\end{align}
With these notations, the phase space coordinates are $X^A=(x,z,P_x,P_z)$. We also rescale the potential as
\begin{align}\label{eq:attractor_dimensionless_potential}
  \mathcal{U}(x,z) &= \cfrac{1}{m_0^2\mP^2}\,\mathcal{V}(\mP x,\mP z) = \cfrac34(1-y)^2+y^2\cfrac{V(\mP z)}{m_0^2\mP^2}\, .
\end{align}
This definition makes the background equations dimensionless. Subscripts on $\mathcal{U}$ denote partial derivatives with respect to $x$ and $z$.

Using $d/dN=H^{-1}d/dt$, the first Hubble slow-roll parameter introduced in \eqref{eq:attractor_end_of_inflation} becomes
\begin{align}\label{eq:attractor_epsilon_definition}
  \epsilon = \cfrac12\left(P_x^2+y\,P_z^2\right)\, ,
\end{align}
and the Friedmann constraint takes the form
\begin{align}\label{eq:attractor_Hubble_constraint_dimensionless}
  H^2 = \cfrac{m_0^2\,\mathcal{U}}{3-\epsilon}\, .
\end{align}
The reduced background equations form the following four-dimensional first-order system:
\begin{align}\label{eq:attractor_background_system}
  \begin{split}
    x' &= P_x\, , \\
    z' &= P_z\, , \\
    P_x' &= -(3-\epsilon)P_x-(3-\epsilon)\cfrac{\mathcal{U}_x}{\mathcal{U}}- \cfrac{\alpha}{2}\,y\,P_z^2\, , \\
    P_z' &= -(3-\epsilon-\alpha P_x)P_z-(3-\epsilon)\cfrac{1}{y}\cfrac{\mathcal{U}_z}{\mathcal{U}}\, .
  \end{split}
\end{align}
Here and below in this section, a prime denotes $d/dN$.

The system contains the exact Starobinsky branch. We define the surface~$\mathcal{M}_{\star}$ in phase space by
\begin{align}\label{eq:attractor_starobinsky_branch}
  z &= 0\, , & P_z &= 0\, .
\end{align}
On this surface one has $z'=P_z=0$. Moreover, from \eqref{eq:attractor_dimensionless_potential} and $V(0) = V'(0) = 0$,
\begin{align}\label{eq:attractor_U_z_branch}
  \mathcal{U}_z(x,0) &= 0\, .
\end{align}
Therefore $P_z'=0$ as well. Hence $\mathcal{M}_{\star}$ is an exact invariant surface of the four-dimensional background system derived from the reduced action \eqref{eq:attractor_reduced_action} and written explicitly in \eqref{eq:attractor_background_system}.

On $\mathcal{M}_{\star}$ the dimensionless potential becomes
\begin{align}\label{eq:attractor_U_on_branch}
  \mathcal{U}(x,0) &= \cfrac34\left(1-\exp[-\alpha x]\right)^2\, .
\end{align}
The remaining equations are
\begin{align}\label{eq:attractor_branch_equations}
  \begin{split}
    x_{\star}' &= P_{x,\star}\, , \\
    P_{x,\star}' &= -(3-\epsilon_{\star})\left(P_{x,\star}+\cfrac{\partial}{\partial x}\ln\mathcal{U}(x_{\star},0)\right)\, , \\
    \epsilon_{\star} &= \cfrac12 \, P_{x,\star}^2\, .
  \end{split}
\end{align}
These are the usual Starobinsky background equations, so this is the space of Starobinsky model solutions embedded in the model under discussion.

It is now necessary to determine whether this branch is stable. In other words, we shall check if the system, given initial conditions near $\mathcal{M}_\star$ at $N_\text{rem}=68$, ends up closer to $\mathcal{M}_{\star}$ at $N_\text{rem}=60$. We do this in two steps. First, we study the variational equations around $\mathcal{M}_{\star}$. This analytic calculation identifies the local damping mechanism and the parameter controlling the transverse restoring force. Second, we compute the finite-time normal-flow map and integrate the nonlinear four-dimensional system derived from the reduced background action within a finite slow-roll region around the branch. Thus, the analytic calculation gives the stability mechanism, while the finite-time and nonlinear calculations determine whether the mechanism is strong enough to contract the chosen region during the specified relaxation interval.

We begin with the variational problem. Let
\begin{align}\label{eq:attractor_vector_field_definition}
  X^{A\prime} &= F^A(X)
\end{align}
be the vector field defined by \eqref{eq:attractor_background_system}. A nearby homogeneous trajectory can be written as
\begin{align}\label{eq:attractor_variation_definition}
  X^A(N) &= X_{\star}^A(N)+\delta X^A(N)\, , &  X_{\star}^A &= (x_{\star},0,P_{x,\star},0)\, .
\end{align}
Keeping only terms linear in $\delta X^A$ gives the variational system
\begin{align}\label{eq:attractor_full_variational_system}
  \delta X^{A\prime} &= A^A{}_B(N)\delta X^B\, , &  A^A{}_B(N) &=  \left.\cfrac{\partial F^A}{\partial X^B}\right|_{X=X_{\star}(N)}\, .
\end{align}
The matrix $A^A{}_B$ is time-dependent because the Starobinsky branch is an invariant two-dimensional submanifold of the four-dimensional phase space rather than a fixed point. The pair $(\delta x,\delta P_x)$ is tangent to $\mathcal{M}_\star$, whereas $(\delta z,\delta P_z)$ is normal to it and controls attraction towards the branch. These homogeneous variations should not be confused with the cosmological perturbations studied in Section~\ref{section_perturbations}.

Because $\mathcal{U}_z(x,0)=0$ and $\mathcal{U}_{xz}(x,0)=0$, the tangent and normal sectors decouple at linear order. The normal variational system is
\begin{align}\label{eq:attractor_normal_variational_system}
  \begin{split}
    \cfrac{d}{dN}
    \begin{pmatrix}
      \delta z\\
      \delta P_z
    \end{pmatrix}
    &=
    A_{\perp}(N)
    \begin{pmatrix}
      \delta z\\
      \delta P_z
    \end{pmatrix}\, , \\
    A_{\perp}(N)
    &=
    \begin{pmatrix}
      0 & 1\\
      -\omega_\chi^2 & -\gamma_\chi
    \end{pmatrix}\, .
  \end{split}
\end{align}
Equivalently, one can obtain a single second-order equation
\begin{align}\label{eq:attractor_normal_second_order}
  \delta z''+\gamma_\chi(N)\delta z' +\omega_\chi^2(N)\delta z &= 0\, .
\end{align}
The coefficients are
\begin{align}\label{eq:attractor_normal_coefficients}
  \gamma_\chi &= 3-\epsilon_{\star}-\alpha P_{x,\star}\, , &
  \omega_\chi^2 &= \cfrac43(3-\epsilon_{\star})\,\mu^2\,\cfrac{y_{\star}}{(1-y_{\star})^2}\, , &
  \mu &\overset{\text{def}}{=} \cfrac{m_\chi}{m_0}\, .
\end{align}

The normal equation \eqref{eq:attractor_normal_second_order} has the form of a damped oscillator with time-dependent coefficients. The term $\gamma_\chi\delta z'$ is the damping term. The term $\omega_\chi^2\delta z$ is the restoring term. For $\omega_\chi^2>0$, it pushes a positive displacement $\delta z$ back toward the branch and similarly pushes a negative displacement toward zero. During Starobinsky slow roll, $P_{x,\star}<0$ and $\epsilon_{\star}<1$, so $\gamma_\chi>0$. The restoring term is also positive for $m_\chi^2>0$, but its magnitude is proportional to $\mu^2$. This is the key analytic point. Light transverse fields have a weak restoring force and relax slowly.

The inequalities $\gamma_\chi>0$ and $\omega_\chi^2>0$ identify the local damping mechanism, but they are not by themselves a complete finite-time contraction statement. There are two reasons. First, the coefficients in \eqref{eq:attractor_normal_second_order} depend on $N$ through the evolving Starobinsky background. Second, contraction must be specified with respect to a chosen norm and over a chosen finite interval. A damped oscillator may have decreasing energy while a particular coordinate norm decreases only weakly, or even temporarily grows, for some initial directions. Therefore, after identifying the analytic damping mechanism, we compute the finite-time flow map of the normal linear system.

We denote the fundamental solution matrix, equivalently the linear transition matrix, of \eqref{eq:attractor_normal_variational_system} by $\Psi_{\perp}$. It maps an initial normal displacement at $N_0$ to the corresponding displacement at $N$ and obeys
\begin{align}\label{eq:attractor_fundamental_matrix}
  \begin{split}
    \begin{pmatrix}
      \delta z(N)\\
      \delta P_z(N)
    \end{pmatrix}
    &=
    \Psi_{\perp}(N,N_0)
    \begin{pmatrix}
      \delta z(N_0)\\
      \delta P_z(N_0)
    \end{pmatrix}\, , \\
    \Psi_{\perp}'(N,N_0)
    &= A_{\perp}(N)\Psi_{\perp}(N,N_0)\, , \\
    \Psi_{\perp}(N_0,N_0)
    &= 1\, .
  \end{split}
\end{align}
Here $1$ denotes the identity matrix in the two-dimensional normal sector.

The following construction is the standard singular-value formulation of finite-time Lyapunov analysis for tangent dynamics \cite{Ott:2002chaos,PikovskyPoliti:2016,Mease:2008ftle}. Introducing the normal displacement vector
\begin{align}\label{eq:attractor_normal_displacement_vector}
  \delta Y_\perp(N) \overset{\text{def}}{=}
  \begin{pmatrix}
    \delta z(N)\\
    \delta P_z(N)
  \end{pmatrix},
\end{align}
the largest singular value of the finite-time map gives the Euclidean operator-norm bound
\begin{align}\label{eq:attractor_singular_value_bound}
  \left\|\delta Y_\perp(N_1)\right\| \leq \sigma_{\max}\!\left[\Psi_\perp(N_1,N_0)\right] \left\|\delta Y_\perp(N_0)\right\| .
\end{align}
We define the largest finite-time transverse Lyapunov exponent by
\begin{align}\label{eq:attractor_finite_time_exponent}
  \lambda_{\max}^{\perp} &\overset{\text{def}}{=} \cfrac{1}{N_1-N_0} \ln\sigma_{\max}\!\left[\Psi_{\perp}(N_1,N_0)\right] .
\end{align}
Equivalently,
\begin{align}\label{eq:attractor_singular_value_growth_relation}
  \sigma_{\max}\!\left[\Psi_{\perp}(N_1,N_0)\right] = \exp\left[\lambda_{\max}^{\perp}(N_1-N_0)\right] .
\end{align}
The quantity $\lambda_{\max}^{\perp}$ is the largest finite-time transverse Lyapunov exponent on $[N_0,N_1]$. It is determined by the largest singular value of the transverse tangent propagator and is not an asymptotic Lyapunov exponent. Thus, $\lambda_{\max}^{\perp}<0$ means that even the infinitesimal direction with the largest final Euclidean norm is contracted over the relaxation interval, whereas $\lambda_{\max}^{\perp}>0$ means that at least one infinitesimal normal direction is stretched. This finite-time criterion is stronger than the mere positivity of the damping and restoring terms because it refers to the complete linear flow map on the chosen interval.

The fundamental matrix tests infinitesimal deviations. We also perform a nonlinear finite-amplitude scan of the four-dimensional system derived from the reduced background action, Eq.~\eqref{eq:attractor_background_system}. For that purpose, we use the transverse distance
\begin{align}\label{eq:attractor_transverse_distance}
  d_{\perp}^2(N) &\overset{\text{def}}{=} \exp[-\alpha x(N)] \left[z(N)^2+P_z(N)^2\right] \, .
\end{align}
This definition includes the same metric factor that appears in the kinetic term of $z$. It is deliberately distinct from the Euclidean norm used for the linear singular-value diagnostic. Therefore, $\lambda_{\max}^{\perp}$ and $R_{\perp}$ are complementary tests of infinitesimal coordinate-space contraction and finite-amplitude kinetic-metric-weighted contraction, respectively. We then define the finite-time contraction ratio
\begin{align}\label{eq:attractor_transverse_ratio}
  R_{\perp} & \overset{\text{def}}{=} \cfrac{d_{\perp}(N_1)}{d_{\perp}(N_0)} \, .
\end{align}
For a given initial condition, $R_{\perp}<1$ means that the trajectory is closer to the Starobinsky branch at the beginning of the physical $60$ e-folds than it was at the beginning of the relaxation interval. The exact branch point $z(N_0)=P_z(N_0)=0$ is excluded from the ratio statistics because it gives the indeterminate ratio $0/0$.

We now specify the finite slow-roll region used for the nonlinear verification. The adiabatic initial data are fixed on the Starobinsky branch at $N_0$, the beginning of the relaxation interval. The transverse initial data are varied in the square
\begin{align}\label{eq:attractor_scan_region}
  z(N_0) &\in [-0.1,0.1]\, , & P_z(N_0) &\in [-0.1,0.1] \, .
\end{align}
This range is inside the slow-roll area. Indeed, the transverse kinetic contribution to $\epsilon$ is
\begin{align}\label{eq:attractor_epsilon_P_z}
  \epsilon_{P_z} &= \cfrac12 \, y\,P_z^2\, .
\end{align}
On the Starobinsky branch, $y\simeq 10^{-2}$ during the interval considered here. Therefore $|P_z(N_0)|\leq0.1$ gives
\begin{align}\label{eq:attractor_epsilon_P_z_estimate}
  \epsilon_{P_z} &\lesssim 6\times10^{-5}\, ,
\end{align}
which is well inside the slow-roll regime. The bounds on $z(N_0)$ coming from the additional potential energy and from the transverse potential-gradient contribution are weaker over the mass range considered here. For instance, near the branch one finds
\begin{align}\label{eq:attractor_z_slow_roll_estimates}
  \cfrac{\mathcal{U}(x,z)-\mathcal{U}(x,0)}{\mathcal{U}(x,0)} &\simeq \cfrac23\,\mu^2\cfrac{y^2}{(1-y)^2}z^2\, .
\end{align}
For $\mu\leq2$, $y\simeq10^{-2}$, and $|z(N_0)|\leq0.1$, this correction is much smaller than unity. In practice, the chosen bound on $z(N_0)$ is a conservative restriction to the field range in which the one-loop EFT and its numerical evaluation remain controlled. The numerical domain used below is summarised in Table~\ref{tab:attractor_scan_domain}.
\begin{table}[t]
  \centering
  \renewcommand{\arraystretch}{1.15}
  \setlength{\tabcolsep}{8pt}
  \begin{tabular}{lc}
    \hline
    Quantity & Value \\
    \hline
    Exact-branch $N_\text{rem}(N_0)$ & $68$ \\
    Pivot condition & $N_\text{rem}(N_1)=60$ \\
    Reference relaxation interval & $8$ e-folds \\
    $z(N_0)$ range & $[-0.1,0.1]$ \\
    $P_z(N_0)$ range & $[-0.1,0.1]$ \\
    Grid & $7\times7$ \\
    $\mu=m_\chi/m_0$ & $0.90,\,0.95,\,1.00,\,1.05,\,1.10$ \\
    \hline
  \end{tabular}
  \caption{Domain of the finite-time stability scan. The initial adiabatic variables are fixed to the common exact-branch state calibrated to $68$ remaining e-folds, while the transverse variables are scanned over the displayed square. The pivot is located separately on each trajectory by $N_\text{rem}=60$.}
  \label{tab:attractor_scan_domain}
\end{table}

The numerical implementation evaluates the explicitly real representation defined by \eqref{the_effective_potential} and \eqref{eq:real_one_loop_potential_continuation}. For each value of $\mu$, we integrate $7\times7=49$ nonlinear trajectories of the reduced four-dimensional background system in $(x,z,P_x,P_z)$, giving $245$ two-field trajectories in total, together with a separate Starobinsky reference. Every displayed curve is a projection of a four-dimensional solution of these reduced background equations. All trajectories reached the end-of-inflation event successfully and remained inside the prescribed field domain.

Table~\ref{tab:attractor_mass_ratio_scan} gives the largest finite-time transverse Lyapunov exponent obtained over the grid, the maximum and median nonlinear contraction ratios, and the number of nonzero grid points with $R_\perp\geq1$.
\begin{table}[t]
  \centering
  \renewcommand{\arraystretch}{1.15}
  \setlength{\tabcolsep}{5pt}
  \begin{tabular}{ccccc}
    \hline
    $\mu$ & $\max\lambda_{\max}^{\perp}$ & $\max R_{\perp}$ & $\operatorname{median}R_{\perp}$ & $R_\perp\geq1$ \\
    \hline
    $0.90$ & $-0.005261$ & $1.018352$ & $0.720086$ & $4/48$ \\
    $0.95$ & $-0.006611$ & $1.007409$ & $0.712348$ & $2/48$ \\
    $1.00$ & $-0.008036$ & $0.995993$ & $0.704276$ & $0/48$ \\
    $1.05$ & $-0.009535$ & $0.984124$ & $0.695883$ & $0/48$ \\
    $1.10$ & $-0.011107$ & $0.971817$ & $0.687182$ & $0/48$ \\
    \hline
  \end{tabular}
  \caption{Finite-time transverse diagnostics over the sampled slow-roll region. The exact branch point is excluded from the nonlinear ratio statistics because it gives $0/0$. The linear transverse map contracts for all five masses, whereas uniform finite-amplitude contraction of the entire sampled square is obtained only for $\mu\geq1$.}
  \label{tab:attractor_mass_ratio_scan}
\end{table}

The largest finite-time transverse Lyapunov exponent is negative for every tested mass, so the infinitesimal normal map contracts over the reference interval. The finite-amplitude criterion is more restrictive. For $\mu\geq1$, every nonzero grid point satisfies $R_\perp<1$, and the entire sampled square is mapped into a smaller region. For $\mu=0.90$ and $0.95$, four and two initial points, respectively, have $R_\perp\geq1$, although the median trajectory contracts. This difference shows why the linear tangent-map criterion and the nonlinear finite-amplitude scan are complementary.

\begin{figure}[t]
  \centering
  \includegraphics[width=\textwidth]{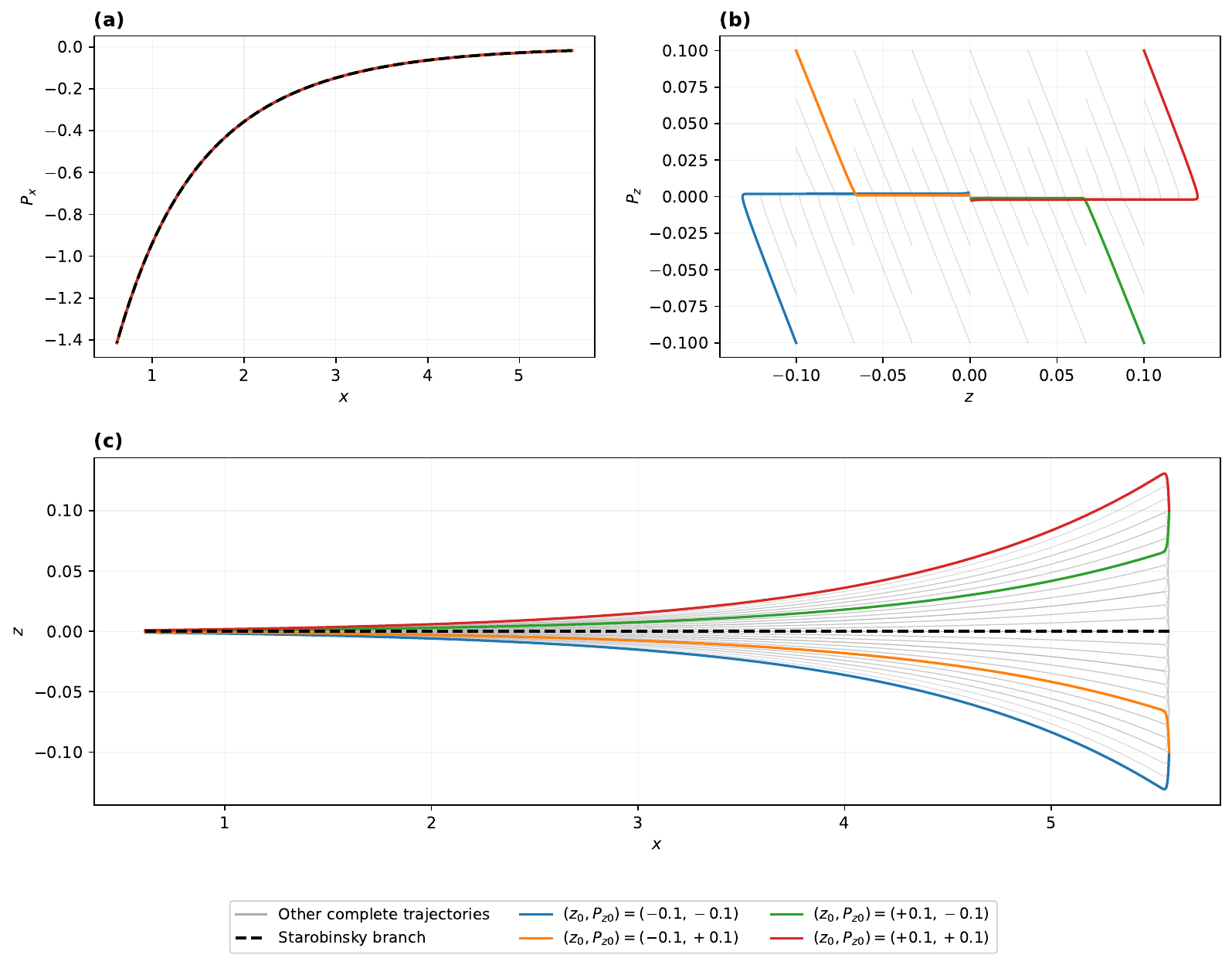}
  \caption{Projections of four-dimensional trajectories of the reduced background system for $\mu=1$ onto the $(x,P_x)$, $(z,P_z)$, and $(x,z)$ planes. The thin grey curves represent the entire $7\times7$ ensemble of two-field trajectories, the coloured curves highlight the four corner initial conditions, and the dashed curve represents the exact Starobinsky branch. All four background variables are integrated before projection.}
  \label{fig:attractor_full_trajectory_projections}
\end{figure}

\begin{figure}[t]
  \centering
  \includegraphics[width=0.72\textwidth]{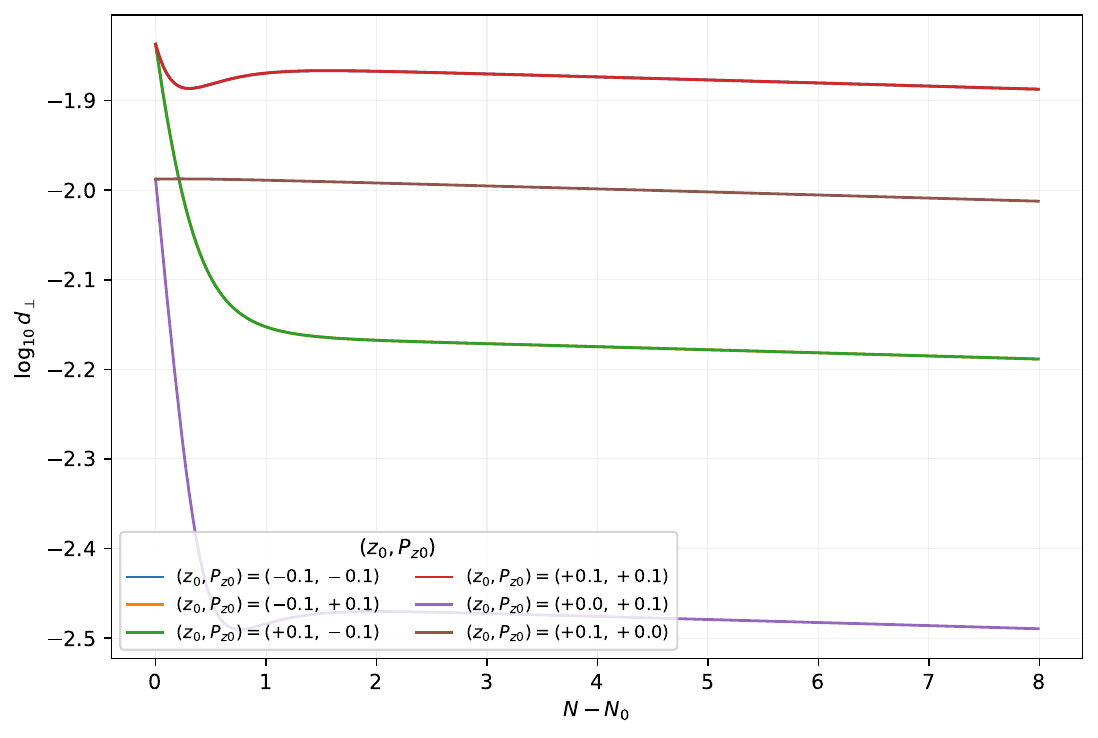}
  \caption{Evolution of the transverse distance $d_\perp$ for representative $\mu=1$ trajectories during the first eight e-folds after the common initial surface.}
  \label{fig:attractor_distance_decay}
\end{figure}

\begin{figure}[t]
  \centering
  \includegraphics[width=0.72\textwidth]{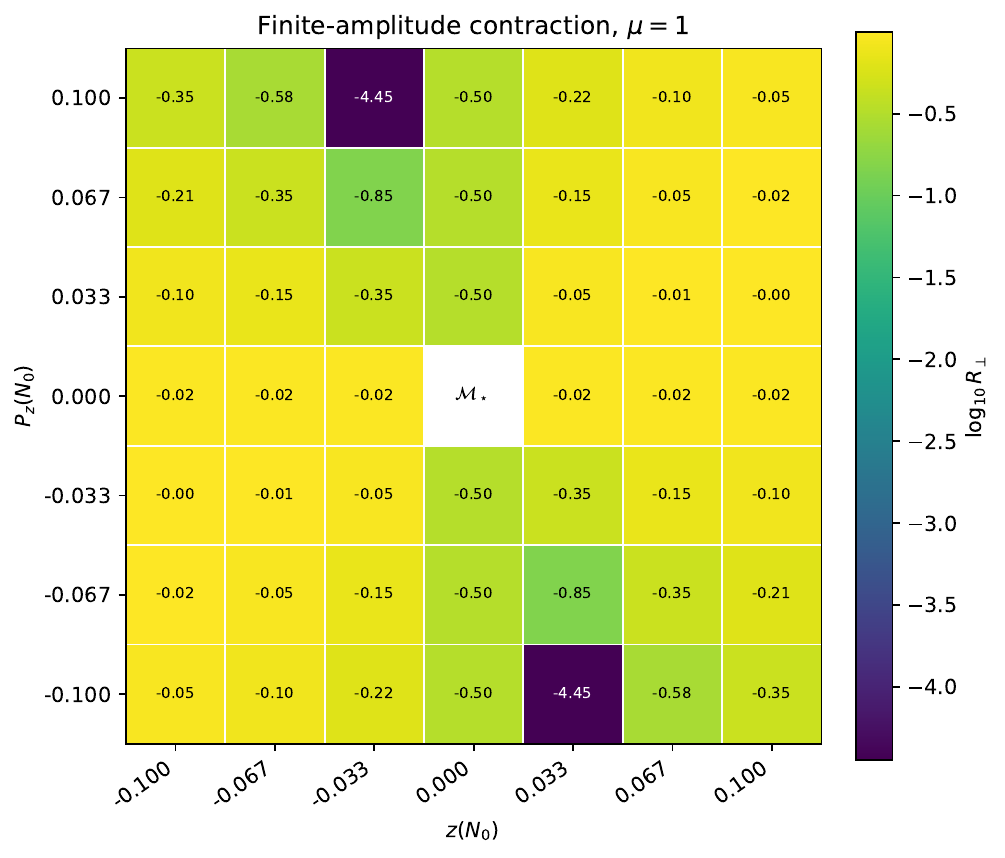}
  \caption{Finite-amplitude scan for $\mu=1$. The colour shows $\log_{10}R_\perp$ over the $7\times7$ grid of transverse initial data. Negative values correspond to contraction.}
  \label{fig:attractor_region_scan}
\end{figure}

Figures~\ref{fig:attractor_full_trajectory_projections}--\ref{fig:attractor_region_scan} illustrate the stable benchmark case $\mu=1$. Figure~\ref{fig:attractor_full_trajectory_projections} shows projections of four-dimensional trajectories of the reduced background system, Figure~\ref{fig:attractor_distance_decay} shows the transverse-distance evolution for representative initial data, and Figure~\ref{fig:attractor_region_scan} displays the finite-amplitude contraction ratio over the full grid.

We therefore arrive at the following conclusion. The reduced two-field system contains the exact Starobinsky branch. The analytic normal equation identifies a damped transverse dynamics, with a restoring term controlled by $\mu^2$. The finite-time linear map contracts for all tested masses, while the nonlinear scan establishes uniform contraction of the entire sampled square only for $\mu\geq1$. For $\mu=0.90$ and $0.95$, the same interval is not sufficient to contract every finite-amplitude direction, although the median trajectory still contracts. This is a local finite-time stability statement for the branch and for the specified $7\times7$ slow-roll sample. It is not a global attractor theorem for the full phase space.

In the next section, we use this branch-restricted background as the basis for the perturbation analysis. The role of the derivative coupling terms in the quadratic action and the possible sourcing of curvature perturbations by entropy modes are treated there separately.

\section{Power spectra of perturbations}\label{section_perturbations}

In the previous section, we showed that the exact Starobinsky branch is a finite-time attractor over the sampled domain for $\mu\geq1$. We now ask whether perturbations generated on this branch can distinguish the radiative two-field completion from the pure Starobinsky model. This requires three separate steps. First, we find the quadratic action for tensor perturbations. Second, we find the quadratic action for scalar perturbations, including the terms generated by the lapse and shift constraints. Finally, we evolve the reduced adiabatic--entropy system and compare the resulting spectra with the Starobinsky model.

The part of the action in the auxiliary field parametrisation \eqref{the_auxiliary_action} quadratic in transverse-traceless perturbations has the form \cite{Kobayashi:2013ina}
\begin{align}\label{eq:perturbation_tensor_action}
  S_T^{(2)} = \cfrac{1}{8}\int dt\,d^3x\,a^3\left[G_T\dot h_{ij}\dot h_{ij}-\cfrac{F_T}{a^2} \, \partial_kh_{ij}\partial_kh_{ij}\right] ,
\end{align}
where the coefficients specifying our model are
\begin{align}\label{eq:perturbation_tensor_coefficients}
  G_T &= \mP^2\left(1+\cfrac{\mu^2\varphi}{3m_\chi^2}\right)+\cfrac{2\beta}{\mP^2}\dot\chi^2\, , &
  F_T &= \mP^2\left(1+\cfrac{\mu^2\varphi}{3m_\chi^2}\right)-\cfrac{2\beta}{\mP^2}\dot\chi^2\, , &
  c_T^2 &= \cfrac{F_T}{G_T}\, .
\end{align}
The tensor sector is free of ghost and gradient instabilities when $G_T>0$ and $F_T>0$. Both coefficients reduce to $\mP^2(1+\mu^2\varphi/(3m_\chi^2))$ on the exact Starobinsky branch. The relative correction to the propagation speed is
\begin{align}\label{eq:perturbation_tensor_relative_correction}
  c_T^2 &= \cfrac{1-\delta_T}{1+\delta_T}\, , &
  \delta_T &\overset{\text{def}}{=} \cfrac{2\beta\dot\chi^2}{\mP^4\left(1+\mu^2\varphi/(3m_\chi^2)\right)}\, .
\end{align}
To establish the characteristic values of $G_T$, $F_T$, and $c_T$, we scanned the entire region discussed in the previous section.

More precisely, for the background solution from the scanned region, we evaluated $G_T$, $F_T$, and $c_T$, since they depend only on the background quantities. In contrast with the previous section, one must specify the value of $\beta$ since it enters the discussed expressions and its influence on the perturbations cannot be neglected. We focused on interval $|\beta|\leq1$ and scanned over the following values
\begin{align}\label{eq:perturbation_beta_test_values}
  \beta \in \{-1,-0.1,0,0.1,1\} \,.
\end{align}
The scan gave the following characteristic values
\begin{align}\label{eq:perturbation_tensor_numerical_bound}
  \min G_T &= 1.651776\,\mP^2\, , & \min F_T &= 1.651776\,\mP^2\, , & \max|c_T^2-1| &<9.8\times10^{-13}
\end{align}
This result shows that in the discussed region of the phase space $G_T$ and $F_T$ do not vanish, but their values remain extremely close, so the overall difference between the speed of light and the speed of tensor perturbations is negligible.

We now consider scalar perturbations in spatially flat slicing. We denote perturbations of $\varphi$ and $\chi$ as $\delta\varphi$ and $\delta\chi$ correspondingly. After eliminating the nondynamical lapse and scalar shift, the constraint-reduced scalar quadratic action reads
\begin{align}\label{eq:perturbation_complete_scalar_action}
  \begin{split}
    S_S^{(2)} = \cfrac12\int dt\,d^3x\,a^3\Bigg[& \begin{pmatrix}\delta\dot\varphi&\delta\dot\chi\end{pmatrix} K \begin{pmatrix}\delta\dot\varphi\\\delta\dot\chi\end{pmatrix} - \cfrac{1}{a^2} \begin{pmatrix} \partial_i\delta\varphi&\partial_i\delta\chi\end{pmatrix} D \begin{pmatrix}\partial_i\delta\varphi\\ \partial_i\delta\chi\end{pmatrix}\\
              &-\begin{pmatrix}\delta\varphi&\delta\chi\end{pmatrix} M\begin{pmatrix}\delta\varphi\\ \delta\chi\end{pmatrix} +2\begin{pmatrix}\delta\varphi&\delta\chi\end{pmatrix} \Omega \begin{pmatrix}\delta\dot\varphi\\\delta\dot\chi\end{pmatrix} \Bigg] \, .
  \end{split}
\end{align}
Here, $K$, $D$, $M$, and $\Omega$ are $2\times 2$ matrices describing different contributions to the quadratic action. Since $[\varphi]=2$ and $[\chi]=1$, the corresponding perturbations have different mass dimensions. The entries of the coefficient matrices therefore have component-dependent dimensions. Explicitly,
\begin{align}\label{eq:perturbation_matrix_dimensions}
  [K_{IJ}] = [D_{IJ}] &= \begin{pmatrix}
    -2&-1\\
    -1&0
  \end{pmatrix}\, , &
  [\Omega_{IJ}] &= \begin{pmatrix}
    -1&0\\
    0&1
  \end{pmatrix}\, , &
  [M_{IJ}] &= \begin{pmatrix}
    0&1\\
    1&2
  \end{pmatrix}\, .
\end{align}
All contributions generated by the lapse and shift constraints are included in the matrices below.
The exact kinetic matrix is
\begin{align}\label{eq:perturbation_kinetic_matrix}
  \begin{split}
    K_{\varphi\varphi}
    &= \cfrac{\mu^4\mP^4\left[3\mP^2\left(1+\mu^2\varphi/(3m_\chi^2)\right)H^2+\dot\chi^2/2\right]}
    {18m_\chi^4\left[\mP^2\left(1+\mu^2\varphi/(3m_\chi^2)\right)H+\mu^2\mP^2\dot\varphi/(6m_\chi^2)+6\beta H\dot\chi^2/\mP^2\right]^2}\, , \\
    K_{\varphi\chi}
    &= \cfrac{4\beta\mu^2H\dot\chi\left[3\mP^2\left(1+\mu^2\varphi/(3m_\chi^2)\right)H^2+\dot\chi^2/2\right]}
    {3m_\chi^2\left[\mP^2\left(1+\mu^2\varphi/(3m_\chi^2)\right)H+\mu^2\mP^2\dot\varphi/(6m_\chi^2)+6\beta H\dot\chi^2/\mP^2\right]^2}\\
    &\quad-\cfrac{\mu^2\mP^2\left(1-12\beta H^2/\mP^2\right)\dot\chi}
    {6m_\chi^2\left[\mP^2\left(1+\mu^2\varphi/(3m_\chi^2)\right)H+\mu^2\mP^2\dot\varphi/(6m_\chi^2)+6\beta H\dot\chi^2/\mP^2\right]}\, , \\
    K_{\chi\chi}
    &= 1-\cfrac{12\beta H^2}{\mP^2}
    +\cfrac{32\beta^2H^2\dot\chi^2\left[3\mP^2\left(1+\mu^2\varphi/(3m_\chi^2)\right)H^2+\dot\chi^2/2\right]}
    {\mP^4\left[\mP^2\left(1+\mu^2\varphi/(3m_\chi^2)\right)H+\mu^2\mP^2\dot\varphi/(6m_\chi^2)+6\beta H\dot\chi^2/\mP^2\right]^2}\\
    &\quad-\cfrac{8\beta H\left(1-12\beta H^2/\mP^2\right)\dot\chi^2}
    {\mP^2\left[\mP^2\left(1+\mu^2\varphi/(3m_\chi^2)\right)H+\mu^2\mP^2\dot\varphi/(6m_\chi^2)+6\beta H\dot\chi^2/\mP^2\right]}\, .
  \end{split}
\end{align}
The exact gradient matrix is
\begin{align}\label{eq:perturbation_gradient_matrix}
  \begin{split}
    D_{\varphi\varphi}
    &= \cfrac{\mu^4\mP^4}
    {18m_\chi^4\left[\mP^2\left(1+\mu^2\varphi/(3m_\chi^2)\right)H+\mu^2\mP^2\dot\varphi/(6m_\chi^2)+6\beta H\dot\chi^2/\mP^2\right]^2}\\
    &\quad\times\Bigg\{3H\left[\mP^2\left(1+\cfrac{\mu^2\varphi}{3m_\chi^2}\right)H+\cfrac{\mu^2\mP^2\dot\varphi}{6m_\chi^2}+\cfrac{6\beta H\dot\chi^2}{\mP^2}\right]\\
    &\hspace{22mm}-\cfrac{d}{dt}\left[\mP^2\left(1+\cfrac{\mu^2\varphi}{3m_\chi^2}\right)H+\cfrac{\mu^2\mP^2\dot\varphi}{6m_\chi^2}+\cfrac{6\beta H\dot\chi^2}{\mP^2}\right]\Bigg\}\, , \\
    D_{\varphi\chi}
    &= -\cfrac{\mu^2\left(\mP^2-20\beta H^2\right)\dot\chi}
    {6m_\chi^2\left[\mP^2\left(1+\mu^2\varphi/(3m_\chi^2)\right)H+\mu^2\mP^2\dot\varphi/(6m_\chi^2)+6\beta H\dot\chi^2/\mP^2\right]}\\
    &\quad+\cfrac1a\cfrac{d}{dt}\left[\cfrac{4a\beta\mu^2H\dot\chi}
      {3m_\chi^2\left[\mP^2\left(1+\mu^2\varphi/(3m_\chi^2)\right)H+\mu^2\mP^2\dot\varphi/(6m_\chi^2)+6\beta H\dot\chi^2/\mP^2\right]}\right]\, , \\
    D_{\chi\chi}
    &= 1-\cfrac{4\beta}{\mP^2}\left(3H^2+2\dot H\right)
    -\cfrac{8\beta H\left(1-12\beta H^2/\mP^2\right)\dot\chi^2}
    {\mP^2\left[\mP^2\left(1+\mu^2\varphi/(3m_\chi^2)\right)H+\mu^2\mP^2\dot\varphi/(6m_\chi^2)+6\beta H\dot\chi^2/\mP^2\right]}\\
    &\quad+\cfrac1a\cfrac{d}{dt}\left[\cfrac{32a\beta^2H^2\dot\chi^2}
      {\mP^4\left[\mP^2\left(1+\mu^2\varphi/(3m_\chi^2)\right)H+\mu^2\mP^2\dot\varphi/(6m_\chi^2)+6\beta H\dot\chi^2/\mP^2\right]}\right]\, .
  \end{split}
\end{align}
The time derivatives in \eqref{eq:perturbation_gradient_matrix} are evaluated with the exact homogeneous equations. The derivative-mixing matrix reads
\begin{align}\label{eq:perturbation_mixing_matrix}
  \begin{split}
    \Omega_{\varphi\varphi}
    &= -\cfrac{\mu^4\mP^4\left(\varphi-6H^2\right)}
          {36m_\chi^4\left[\mP^2\left(1+\mu^2\varphi/(3m_\chi^2)\right)H+\mu^2\mP^2\dot\varphi/(6m_\chi^2)+6\beta H\dot\chi^2/\mP^2\right]}\\
          &\quad-\cfrac{\mu^4\mP^4H\left[\dot\chi^2/2-\mu^2\mP^2H\dot\varphi/(2m_\chi^2)-18\beta H^2\dot\chi^2/\mP^2\right]}
          {18m_\chi^4\left[\mP^2\left(1+\mu^2\varphi/(3m_\chi^2)\right)H+\mu^2\mP^2\dot\varphi/(6m_\chi^2)+6\beta H\dot\chi^2/\mP^2\right]^2}\, , \\
          \Omega_{\varphi\chi}
          &= -\cfrac{2\beta\mu^2H\dot\chi\left(\varphi-6H^2\right)}
                {3m_\chi^2\left[\mP^2\left(1+\mu^2\varphi/(3m_\chi^2)\right)H+\mu^2\mP^2\dot\varphi/(6m_\chi^2)+6\beta H\dot\chi^2/\mP^2\right]}\\
                &\quad-\cfrac{4\beta\mu^2H^2\dot\chi\left[\dot\chi^2/2-\mu^2\mP^2H\dot\varphi/(2m_\chi^2)-18\beta H^2\dot\chi^2/\mP^2\right]}
                {3m_\chi^2\left[\mP^2\left(1+\mu^2\varphi/(3m_\chi^2)\right)H+\mu^2\mP^2\dot\varphi/(6m_\chi^2)+6\beta H\dot\chi^2/\mP^2\right]^2}\\
                &\quad+\cfrac{\mu^2\mP^2H\left(1-12\beta H^2/\mP^2\right)\dot\chi}
                {6m_\chi^2\left[\mP^2\left(1+\mu^2\varphi/(3m_\chi^2)\right)H+\mu^2\mP^2\dot\varphi/(6m_\chi^2)+6\beta H\dot\chi^2/\mP^2\right]}\, , \\
                \Omega_{\chi\varphi}
                &= -\cfrac{\mu^2\mP^2V_{,\chi}}
                      {6m_\chi^2\left[\mP^2\left(1+\mu^2\varphi/(3m_\chi^2)\right)H+\mu^2\mP^2\dot\varphi/(6m_\chi^2)+6\beta H\dot\chi^2/\mP^2\right]}\\
                      &\quad+\cfrac{\mu^2\mP^2\left(1-12\beta H^2/\mP^2\right)\dot\chi
                        \left[\dot\chi^2/2-\mu^2\mP^2H\dot\varphi/(2m_\chi^2)-18\beta H^2\dot\chi^2/\mP^2\right]}
                      {6m_\chi^2\left[\mP^2\left(1+\mu^2\varphi/(3m_\chi^2)\right)H+\mu^2\mP^2\dot\varphi/(6m_\chi^2)+6\beta H\dot\chi^2/\mP^2\right]^2}\, , \\
                      \Omega_{\chi\chi}
                      &= -\cfrac{4\beta H\dot\chi V_{,\chi}}
                            {\mP^2\left[\mP^2\left(1+\mu^2\varphi/(3m_\chi^2)\right)H+\mu^2\mP^2\dot\varphi/(6m_\chi^2)+6\beta H\dot\chi^2/\mP^2\right]}\\
                            &\quad+\cfrac{4\beta H\left(1-12\beta H^2/\mP^2\right)\dot\chi^2
                              \left[\dot\chi^2/2-\mu^2\mP^2H\dot\varphi/(2m_\chi^2)-18\beta H^2\dot\chi^2/\mP^2\right]}
                            {\mP^2\left[\mP^2\left(1+\mu^2\varphi/(3m_\chi^2)\right)H+\mu^2\mP^2\dot\varphi/(6m_\chi^2)+6\beta H\dot\chi^2/\mP^2\right]^2}\\
                            &\quad-\cfrac{\left(1-12\beta H^2/\mP^2\right)^2\dot\chi^2}
                            {2\left[\mP^2\left(1+\mu^2\varphi/(3m_\chi^2)\right)H+\mu^2\mP^2\dot\varphi/(6m_\chi^2)+6\beta H\dot\chi^2/\mP^2\right]}\, .
  \end{split}
\end{align}
The exact mass-sector coefficient matrix is
\begin{align}\label{eq:perturbation_mass_matrix}
  \begin{split}
    M_{\varphi\varphi}
    &= \cfrac{\mu^2\mP^2}{6m_\chi^2}
    -\cfrac{\mu^4\mP^4H^2
      \left[\dot\chi^2/2-3\mP^2\left(1+\mu^2\varphi/(3m_\chi^2)\right)H^2-\mu^2\mP^2H\dot\varphi/m_\chi^2-36\beta H^2\dot\chi^2/\mP^2\right]}
    {18m_\chi^4\left[\mP^2\left(1+\mu^2\varphi/(3m_\chi^2)\right)H+\mu^2\mP^2\dot\varphi/(6m_\chi^2)+6\beta H\dot\chi^2/\mP^2\right]^2}\\
    &\quad-\cfrac{\mu^4\mP^4H\left(\varphi-6H^2\right)}
    {18m_\chi^4\left[\mP^2\left(1+\mu^2\varphi/(3m_\chi^2)\right)H+\mu^2\mP^2\dot\varphi/(6m_\chi^2)+6\beta H\dot\chi^2/\mP^2\right]}\, , \\
    M_{\varphi\chi}
    &= \cfrac{\mu^2\mP^2H\left(1-12\beta H^2/\mP^2\right)\dot\chi
      \left[\dot\chi^2/2-3\mP^2\left(1+\mu^2\varphi/(3m_\chi^2)\right)H^2-\mu^2\mP^2H\dot\varphi/m_\chi^2-36\beta H^2\dot\chi^2/\mP^2\right]}
    {6m_\chi^2\left[\mP^2\left(1+\mu^2\varphi/(3m_\chi^2)\right)H+\mu^2\mP^2\dot\varphi/(6m_\chi^2)+6\beta H\dot\chi^2/\mP^2\right]^2}\\
    &\quad+\cfrac{\mu^2\mP^2\left[\left(\varphi-6H^2\right)\left(1-12\beta H^2/\mP^2\right)\dot\chi-2HV_{,\chi}\right]}
    {12m_\chi^2\left[\mP^2\left(1+\mu^2\varphi/(3m_\chi^2)\right)H+\mu^2\mP^2\dot\varphi/(6m_\chi^2)+6\beta H\dot\chi^2/\mP^2\right]}\, , \\
    M_{\chi\chi}
    &= V_{,\chi\chi}
    -\cfrac{\left(1-12\beta H^2/\mP^2\right)^2\dot\chi^2
      \left[\dot\chi^2/2-3\mP^2\left(1+\mu^2\varphi/(3m_\chi^2)\right)H^2-\mu^2\mP^2H\dot\varphi/m_\chi^2-36\beta H^2\dot\chi^2/\mP^2\right]}
    {2\left[\mP^2\left(1+\mu^2\varphi/(3m_\chi^2)\right)H+\mu^2\mP^2\dot\varphi/(6m_\chi^2)+6\beta H\dot\chi^2/\mP^2\right]^2}\\
    &\quad+\cfrac{\left(1-12\beta H^2/\mP^2\right)\dot\chi V_{,\chi}}
    {\mP^2\left(1+\mu^2\varphi/(3m_\chi^2)\right)H+\mu^2\mP^2\dot\varphi/(6m_\chi^2)+6\beta H\dot\chi^2/\mP^2}\, .
  \end{split}
\end{align}

The symmetric and antisymmetric parts of the derivative-mixing matrix are
\begin{align}\label{eq:perturbation_mixing_decomposition}
  \Omega_{\rm S} &= \cfrac12\left(\Omega+\Omega^{\rm T}\right)\, , & \Omega_{\rm A} &= \cfrac12\left(\Omega-\Omega^{\rm T}\right)\, .
\end{align}
The symmetric part can be removed by integration by parts. The corresponding symmetric mass-sector coefficient matrix is
\begin{align}\label{eq:perturbation_canonical_mass_matrix}
  M_{\rm can} = M+\dot\Omega_{\rm S}+3H\Omega_{\rm S}\, .
\end{align}
The irreducible derivative mixing is contained in $\Omega_{\rm A}$. The eigenvalues of $M_{\rm can}$ are not physical mass-squared eigenvalues before kinetic normalisation.

The determinant of the kinetic matrix is
\begin{align}\label{eq:perturbation_kinetic_determinant}
  \begin{split}
    \det K ={}& \cfrac{\mu^4\mP^4H^2} {6m_\chi^4\left[\mP^2\left(1+\mu^2\varphi/(3m_\chi^2)\right)H+\mu^2\mP^2\dot\varphi/(6m_\chi^2)+6\beta H\dot\chi^2/\mP^2\right]^2} \\
    &\times\left[\mP^2\left(1+\cfrac{\mu^2\varphi}{3m_\chi^2}\right)+\cfrac{2\beta\dot\chi^2}{\mP^2}\right] \left(1-\cfrac{12\beta H^2}{\mP^2}\right) \, .
  \end{split}
\end{align}
Moreover,
\begin{align}\label{eq:perturbation_kinetic_leading_minor}
  K_{\varphi\varphi} = \cfrac{\mu^4\mP^4\left[3\mP^2\left(1+\mu^2\varphi/(3m_\chi^2)\right)H^2+\dot\chi^2/2\right]}{18m_\chi^4\left[\mP^2\left(1+\mu^2\varphi/(3m_\chi^2)\right)H+\mu^2\mP^2\dot\varphi/(6m_\chi^2)+6\beta H\dot\chi^2/\mP^2\right]^2}>0
\end{align}
for $1+\mu^2\varphi/(3m_\chi^2)>0$ on an expanding background. At regular points where the common denominator above does not vanish, the scalar no-ghost conditions are
\begin{align}\label{eq:perturbation_scalar_no_ghost_conditions}
  \mP^2\left(1+\cfrac{\mu^2\varphi}{3m_\chi^2}\right)+\cfrac{2\beta\dot\chi^2}{\mP^2} &>0\, , & 1-\cfrac{12\beta H^2}{\mP^2} &>0\, .
\end{align}
The gradient determinant is
\begin{align}\label{eq:perturbation_gradient_determinant}
  \begin{split}
    \det D ={}& \cfrac{\mu^4\mP^4}{36m_\chi^4\left[\mP^2\left(1+\mu^2\varphi/(3m_\chi^2)\right)H+\mu^2\mP^2\dot\varphi/(6m_\chi^2)+6\beta H\dot\chi^2/\mP^2\right]^2}\\
    &\times\Bigg\{2\left[1-\cfrac{4\beta}{\mP^2}\left(3H^2+2\dot H\right)\right]
    \Bigg[3H\left(\mP^2\left(1+\cfrac{\mu^2\varphi}{3m_\chi^2}\right)H+\cfrac{\mu^2\mP^2\dot\varphi}{6m_\chi^2}+\cfrac{6\beta H\dot\chi^2}{\mP^2}\right)\\
      & \hspace{15pt}-\cfrac{d}{dt}\left(\mP^2\left(1+\cfrac{\mu^2\varphi}{3m_\chi^2}\right)H+\cfrac{\mu^2\mP^2\dot\varphi}{6m_\chi^2}+\cfrac{6\beta H\dot\chi^2}{\mP^2}\right)\Bigg]\\
    &\hspace{30pt}-\Bigg[\cfrac{8\beta H^2\dot\chi}{\mP^2}-\cfrac{d}{dt}\left(\cfrac{8\beta H\dot\chi}{\mP^2}\right)+\left(1-\cfrac{12\beta H^2}{\mP^2}\right)\dot\chi\Bigg]^2\Bigg\} \, .
  \end{split}
\end{align}
Positive definiteness of $D$ leads to the following more complicated condition
\begin{align}\label{eq:perturbation_scalar_gradient_conditions}
  \begin{split}
    &3H\left[\mP^2\left(1+\cfrac{\mu^2\varphi}{3m_\chi^2}\right)H+\cfrac{\mu^2\mP^2\dot\varphi}{6m_\chi^2}+\cfrac{6\beta H\dot\chi^2}{\mP^2}\right] -\cfrac{d}{dt}\left[\mP^2\left(1+\cfrac{\mu^2\varphi}{3m_\chi^2}\right)H+\cfrac{\mu^2\mP^2\dot\varphi}{6m_\chi^2}+\cfrac{6\beta H\dot\chi^2}{\mP^2}\right]>0\, , \\
    &2\left[1-\cfrac{4\beta}{\mP^2}\left(3H^2+2\dot H\right)\right] \Bigg\{3H\left[\mP^2\left(1+\cfrac{\mu^2\varphi}{3m_\chi^2}\right)H+\cfrac{\mu^2\mP^2\dot\varphi}{6m_\chi^2}+\cfrac{6\beta H\dot\chi^2}{\mP^2}\right] \\
    &\hspace{10pt}-\cfrac{d}{dt}\left[\mP^2\left(1+\cfrac{\mu^2\varphi}{3m_\chi^2}\right)H+\cfrac{\mu^2\mP^2\dot\varphi}{6m_\chi^2}+\cfrac{6\beta H\dot\chi^2}{\mP^2}\right]\Bigg\} >\Bigg[\cfrac{8\beta H^2\dot\chi}{\mP^2}-\cfrac{d}{dt}\left(\cfrac{8\beta H\dot\chi}{\mP^2}\right)+\left(1-\cfrac{12\beta H^2}{\mP^2}\right)\dot\chi\Bigg]^2 \, .
  \end{split}
\end{align}
Since $\delta\varphi$ and $\delta\chi$ have different mass dimensions, the numerical characteristic problem is first transformed to the dimensionally homogeneous perturbations $(\delta\phi,\delta\chi)$ induced by the exact scalar redefinition \eqref{eq:model_scalar_field_redefinition}. At linear order,
\begin{align}\label{eq:perturbation_homogeneous_dimension_basis}
  \begin{pmatrix}\delta\varphi\\\delta\chi\end{pmatrix}
  &= J
  \begin{pmatrix}\delta\phi\\\delta\chi\end{pmatrix}\, , &
  J &\overset{\text{def}}{=}
  \begin{pmatrix}
    \sqrt{6}\,m_0^2F(\varphi)/\mP & 0\\
    0 & 1
  \end{pmatrix}.
\end{align}
The principal matrices in this basis are
\begin{align}\label{eq:perturbation_hatted_principal_matrices}
  \overline K &= J^T KJ\, , & \overline D &= J^T DJ\, .
\end{align}
The kinetic matrix must be positive definite. The squared scalar propagation speeds are obtained as the generalised eigenvalues of the kinetic-gradient pair $(\overline K,\overline D)$, or equivalently as the roots of the basis-independent characteristic equation
\begin{align}\label{eq:perturbation_scalar_sound_speeds}
  \det\left(D-c_s^2K\right)=0\, .
\end{align}
The corresponding eigenvectors define the instantaneous principal propagation directions and need not coincide with the field-coordinate perturbations away from the exact branch \cite{Kobayashi:2013ina}.

For the trajectory-wise diagnostics, the dimensionless off-diagonal kinetic and gradient entries are
\begin{align}\label{eq:perturbation_off_diagonal_diagnostics}
  \rho_K &= \cfrac{\overline K_{\phi\chi}}{\sqrt{\overline K_{\phi\phi}\overline K_{\chi\chi}}}\, , &
  \rho_D &= \cfrac{\overline D_{\phi\chi}}{\sqrt{\overline D_{\phi\phi}\overline D_{\chi\chi}}}\, .
\end{align}
To evaluate the derivative-mixing and mass sectors after kinetic normalisation, we choose a time-dependent matrix $S_{\overline K}$ satisfying $S_{\overline K}^T\overline KS_{\overline K}= 1$ and write
\begin{align}\label{eq:perturbation_kinetic_normalisation}
  \begin{pmatrix}\delta\varphi\\\delta\chi\end{pmatrix}
  =J S_{\overline K}
  \begin{pmatrix}v_1\\v_2\end{pmatrix}.
\end{align}
The time dependence of the complete transformation is retained in the quadratic action. After the symmetric derivative mixing is absorbed by integration by parts, we denote the remaining antisymmetric mixing matrix by $\Omega_{\rm phys}$ and the resulting symmetric mass-sector matrix by $M_{\rm phys}$. We report $\|\Omega_{\rm phys}\|/H$ and the eigenvalues of $M_{\rm phys}/H^2$; the raw dimensionful eigenvalues of $M$ are not interpreted as physical masses.

The exact Starobinsky branch is defined by $\chi=0$, $\dot\chi=0$, and $V_{,\chi}(0)=0$. On this branch, the principal matrices are diagonal,
\begin{align}\label{eq:perturbation_starobinsky_branch_principal_matrices}
  K &= \begin{pmatrix}
    \cfrac{\mu^4\mP^6\left(1+\mu^2\varphi/(3m_\chi^2)\right)H^2}
          {6m_\chi^4\left[\mP^2\left(1+\mu^2\varphi/(3m_\chi^2)\right)H+\mu^2\mP^2\dot\varphi/(6m_\chi^2)\right]^2} & 0\\
          0 & 1-\cfrac{12\beta H^2}{\mP^2}
  \end{pmatrix}\, , \nonumber\\
  D &= \begin{pmatrix}
    \cfrac{\mu^4\mP^6\left(1+\mu^2\varphi/(3m_\chi^2)\right)H^2}
          {6m_\chi^4\left[\mP^2\left(1+\mu^2\varphi/(3m_\chi^2)\right)H+\mu^2\mP^2\dot\varphi/(6m_\chi^2)\right]^2} & 0\\
          0 & 1-\cfrac{4\beta}{\mP^2}\left(3H^2+2\dot H\right)
  \end{pmatrix}\, .
\end{align}
The scalaron mode remains luminal. The transverse propagation speed is
\begin{align}\label{eq:perturbation_starobinsky_branch_speeds}
  c_\varphi^2 &=1\, , &
  c_\chi^2 &= \cfrac{1-4\beta\left(3H^2+2\dot H\right)/\mP^2}{1-12\beta H^2/\mP^2} =1-\cfrac{8\beta\dot H/\mP^2}{1-12\beta H^2/\mP^2}\, .
\end{align}
The irreducible derivative mixing and the off-diagonal element of $M_{\rm can}$ also vanish,
\begin{align}\label{eq:perturbation_starobinsky_branch_mixing}
  \Omega_{\rm A} &=0\, , & (M_{\rm can})_{\varphi\chi} &=0\, .
\end{align}
The complete scalar sector is free of ghost and high-frequency gradient instabilities on the exact branch whenever
\begin{align}\label{eq:perturbation_starobinsky_branch_stability}
  1-\cfrac{12\beta H^2}{\mP^2} &>0\, , &  1-\cfrac{4\beta}{\mP^2}\left(3H^2+2\dot H\right) &>0\, .
\end{align}

We evaluate the complete matrices on the $246$ stored backgrounds and on the five values of $\beta$ in \eqref{eq:perturbation_beta_test_values}, giving $1230$ trajectory--coefficient cases. Analytic homogeneous equations are used for the time derivatives entering $D$, and the scalar characteristics are obtained from \eqref{eq:perturbation_scalar_sound_speeds}. The exact regression identity $D(0)=K(0)$ is satisfied with a maximum relative residual $8.3\times10^{-16}$.

The extrema grouped by $\beta$ are summarised in Table~\ref{tab:perturbation_complete_matrix_scan}. Over the entire scan, we obtain
\begin{align}\label{eq:perturbation_complete_matrix_global_bounds}
  \min\lambda(\overline K) &=0.9999824\, , & \max|c_s^2-1| &<4.0\times10^{-10}\, , \nonumber \\
  \max|\rho_K| &<4.2\times10^{-3}\, , & \max|\rho_D| &<4.2\times10^{-3}\, ,\nonumber \\
  \max\cfrac{\|\Omega_{\rm phys}\|}{H} &=8.49\times10^{-3}\, .
\end{align}
Both scalar propagation speeds equal unity to machine precision at $\beta=0$. The lowest kinetically normalised mass eigenvalue reaches approximately $-2.25H^2$, indicating a transient tachyonic mass direction but not a ghost or high-frequency gradient instability. The 1001-point production grid was compared with 501- and 2001-point grids for the exact branch and two extremal trajectories. The sound-speed extrema are unchanged at the displayed precision, while the lowest mass eigenvalue changes by $4.4\times10^{-3}$ between 1001 and 2001 points.

\begin{table}[t]
  \centering
  \renewcommand{\arraystretch}{1.15}
  \setlength{\tabcolsep}{3.5pt}
  \begin{tabular}{ccccccc}
    \hline
    $\beta$ & $\min\lambda(\overline K)$ & $\max|c_s^2-1|$ & $\max|\rho_K|$ & $\max|\rho_D|$ & $\max\|\Omega_{\rm phys}\|/H$ & $\max|c_T^2-1|$ \\
    \hline
    $-1$   & $0.99998246$ & $3.93\times10^{-10}$ & $4.169\times10^{-3}$ & $4.169\times10^{-3}$ & $8.487\times10^{-3}$ & $9.79\times10^{-13}$ \\
    $-0.1$ & $0.99998244$ & $3.93\times10^{-11}$ & $4.169\times10^{-3}$ & $4.169\times10^{-3}$ & $8.487\times10^{-3}$ & $9.79\times10^{-14}$ \\
    $0$    & $0.99998243$ & $8.88\times10^{-16}$ & $4.169\times10^{-3}$ & $4.169\times10^{-3}$ & $8.487\times10^{-3}$ & $0$ \\
    $0.1$  & $0.99998243$ & $3.93\times10^{-11}$ & $4.169\times10^{-3}$ & $4.169\times10^{-3}$ & $8.487\times10^{-3}$ & $9.80\times10^{-14}$ \\
    $1$    & $0.99998241$ & $3.93\times10^{-10}$ & $4.169\times10^{-3}$ & $4.169\times10^{-3}$ & $8.487\times10^{-3}$ & $9.79\times10^{-13}$ \\
    \hline
  \end{tabular}
  \caption{Global extrema of the complete scalar and tensor quadratic-action diagnostics for each tested value of~$\beta$. Every row contains $246$ backgrounds.}
  \label{tab:perturbation_complete_matrix_scan}
\end{table}

These results show that the derivative interaction preserves the positivity of the principal perturbation operator and changes the scalar and tensor characteristic speeds only imperceptibly on the mapped slow-roll backgrounds. The reduced spectral calculation below can therefore be interpreted as an accurate description of the scalar characteristics over the tested branch and coefficient interval, subject to the stated fixed-background comparison.

Having established the stability of the complete quadratic operator, we next quantify the accuracy of using the reduced scalar equations for the production spectra. We close the relation between the complete quadratic operator and the reduced production spectra by two complementary calculations. Both use the same stored trajectories of the four-dimensional background system derived from the reduced slow-roll action. Thus, the comparison isolates the derivative coupling in the \emph{linear perturbation operator}; it does not add the derivative-coupling correction to the homogeneous background. This fixed-background scope is common to the stability scan above and to the validation reported here.

First, for every tested trajectory and for $\beta\in\{-1,0,1\}$, we evolve the complete constraint-reduced scalar system. We denote by $P_{\mathcal R}^{\rm complete}(k;\beta)$ the final dimensionless curvature-perturbation power spectrum obtained from this evolution for wavenumber $k$ and derivative-coupling coefficient $\beta$, on the same fixed reduced background. The pointwise correction is
\begin{align}\label{eq:complete_spectrum_beta_correction}
  \Delta_\beta(k) \equiv \frac{P_{\mathcal R}^{\rm complete}(k;\beta)}{P_{\mathcal R}^{\rm complete}(k;0)}-1.
\end{align}
The denominator is the corresponding complete result with the derivative coupling switched off, $\beta=0$. This comparison measures the derivative-coupling effect without folding in the difference between the complete and reduced implementations. We test the latter separately at $\beta=0$ through
\begin{align}\label{eq:complete_reduced_beta_zero_regression}
  \Delta_0^{\rm reg}(k) \equiv \frac{P_{\mathcal R}^{\rm complete}(k;0)}{P_{\mathcal R}^{\rm reduced}(k)}-1.
\end{align}
Here $P_{\mathcal R}^{\rm reduced}(k)$ denotes the final dimensionless curvature spectrum obtained from the reduced scalar evolution on the same background.

We use the CPU calculation as the reference production result. By ``reference'' we mean that it supplies the primary production estimate and the CPU numerical-convergence allowance: the $51$-mode scan covers all $246$ trajectories, while $501$-mode calculations cover every extremum and the dynamically selected refinement cases, and the matrix grid, solver tolerances, and Bunch--Davies surface are varied independently. This status specifies how the numerical error budget is organised; it does not mean that the GPU result is discarded. The reference CPU calculation gives $\max|\Delta_\beta|=\csname VALIDATION_CPU_OBSERVED_MAX\endcsname$ and $\epsilon_{\rm numerical}=\csname VALIDATION_NUMERICAL_ALLOWANCE\endcsname$.

The dense $501$-mode GPU calculation is an independent full-domain cross-check over all $246$ trajectories. Under the predefined promotion criterion, it would be treated as a second production determination only if the CPU--GPU difference were below one tenth of the CPU numerical allowance. It passes the precision, Wronskian, speed, memory, completeness, and physical-bound checks, but not this deliberately stringent cross-implementation criterion. We therefore do not average it with the CPU result as a second production estimate. Instead, the conservative bound uses its larger observed correction, $\csname VALIDATION_OBSERVED_MAX\endcsname$, and includes the entire CPU--GPU discrepancy, $\epsilon_{\rm CPU\text{-}GPU}=\csname VALIDATION_CPU_GPU_ALLOWANCE\endcsname$, as an additional allowance. Thus,
\begin{align}\label{eq:complete_spectrum_validation_bound}
  \max_{\rm scan}|\Delta_\beta|+\epsilon_{\rm numerical}
  +\epsilon_{\rm CPU\text{-}GPU}
  \leq \csname VALIDATION_COMPLETE_BOUND\endcsname <10^{-3}.
\end{align}
The independent $\beta=0$ regression is $\max|\Delta_0^{\rm reg}|=\csname VALIDATION_BETA_ZERO_REGRESSION\endcsname$, and the largest Wronskian residual over the two implementations is $\csname VALIDATION_WRONSKIAN\endcsname$.

Second, we express every $\beta\ne0$ operator in the kinetic-normalised basis constructed at $\beta=0$ on the same background. Denoting matrices in this common basis by hats, we evaluate the spectral norms
\begin{align}\label{eq:canonical_coefficient_validation_norms}
 \|\Delta\widehat K\|_2,\qquad
 \|\Delta\widehat D\|_2,\qquad
 \frac{\|\Delta\Omega_{\rm phys}\|_2}{H},\qquad
 \frac{\|\Delta M_{\rm phys}\|_2}{H^2},
\end{align}
together with the changes in the canonical mass eigenvalues, sound speeds, mixing norm, and the $N$-integrated mass and mixing differences. The largest mass- and mixing-sector differences are $\csname VALIDATION_DELTA_M\endcsname$ and $\csname VALIDATION_DELTA_OMEGA\endcsname$, respectively. Their extrema are finite, explicitly located in trajectory, $\beta$, and e-fold time, and stable under refinement from $2001$ to $4001$ coefficient points. The two routes therefore test both the accumulated spectral effect and the instantaneous operator deformation.

\begin{table}[t]
  \centering
  \renewcommand{\arraystretch}{1.13}
  \setlength{\tabcolsep}{6pt}
  \begin{tabular}{ll}
    \hline
    Validation quantity & Global bound or maximum \\
    \hline
    Reference CPU correction $\max|\Delta_\beta|$ & $\csname VALIDATION_CPU_OBSERVED_MAX\endcsname$ \\
    Full-domain GPU correction (supplementary) & $\csname VALIDATION_OBSERVED_MAX\endcsname$ \\
    Numerical allowance $\epsilon_{\rm numerical}$ & $\csname VALIDATION_NUMERICAL_ALLOWANCE\endcsname$ \\
    CPU--GPU allowance $\epsilon_{\rm CPU\text{-}GPU}$ & $\csname VALIDATION_CPU_GPU_ALLOWANCE\endcsname$ \\
    Conservative spectrum bound & $\csname VALIDATION_COMPLETE_BOUND\endcsname$ \\
    $\beta=0$ complete--reduced regression & $\csname VALIDATION_BETA_ZERO_REGRESSION\endcsname$ \\
    Wronskian residual & $\csname VALIDATION_WRONSKIAN\endcsname$ \\
    $\max\|\Delta\widehat K\|_2$ & $\csname VALIDATION_DELTA_K\endcsname$ \\
    $\max\|\Delta\widehat D\|_2$ & $\csname VALIDATION_DELTA_D\endcsname$ \\
    $\max\|\Delta\Omega_{\rm phys}\|_2/H$ & $\csname VALIDATION_DELTA_OMEGA\endcsname$ \\
    $\max\|\Delta M_{\rm phys}\|_2/H^2$ & $\csname VALIDATION_DELTA_M\endcsname$ \\
    \hline
  \end{tabular}
  \caption{Direct spectral and common-basis coefficient validations of the omitted derivative coupling over the tested fixed-background domain. The CPU calculation supplies the reference production estimate and its convergence allowance. The dense GPU calculation is an independent supplementary cross-check; its larger observed correction and the full CPU--GPU discrepancy are included in the conservative bound.}
  \label{tab:derivative_coupling_validation}
\end{table}

\begin{figure}[t]
  \centering
  \includegraphics[width=0.98\textwidth]{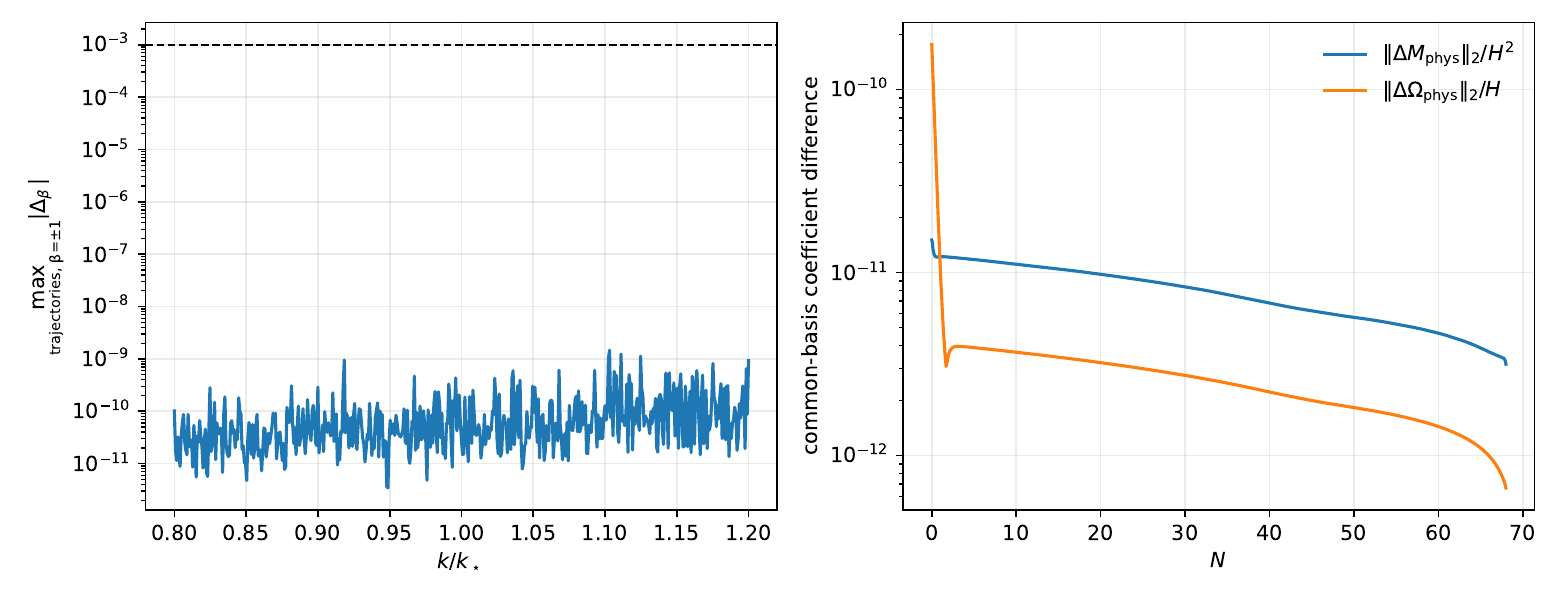}
  \caption{Two complementary validations of the reduced spectral evolution. Left: the supplementary full-domain GPU envelope of the complete-spectrum correction for $\beta=\pm1$; the dashed line marks the $10^{-3}$ physical tolerance. Right: the e-fold-dependent common-basis envelopes of the canonically normalised mass and derivative-mixing differences relative to $\beta=0$.}
  \label{fig:derivative_coupling_validation}
\end{figure}

Since the conservative bound in Eq.~\eqref{eq:complete_spectrum_validation_bound} is well below the accuracy relevant here, we omit the corresponding non-minimal coupling terms from the production scalar evolution and proceed with the reduced action
\begin{align}\label{eq:section5_reduced_scalar_action}
  S_\text{red} = \int d^4x\sqrt{-g} \left[ \cfrac{\mP^2}{2}R - \cfrac12(\nabla\phi)^2 - \cfrac12 \, e^{2b(\phi)}(\nabla\chi)^2 - \mathcal{V}(\phi,\chi) \right],
\end{align}
with
\begin{align}\label{eq:section5_reduced_scalar_identification}
  b(\phi) &= -\cfrac{\phi}{\sqrt{6}\,\mP}, & \mathcal{V}(\phi,\chi) &= U(\phi)+\exp\left[-2\sqrt{\cfrac23}\,\cfrac{\phi}{\mP}\right]V(\chi).
\end{align}
This is the two-field system to which the adiabatic-entropy formalism of \cite{Lalak:2007vi} applies.

We derive the scalar perturbation equations by expanding the metric and the scalar fields around the homogeneous cosmological background. It is convenient for this derivation to work in the longitudinal gauge,
\begin{align}\label{eq:section5_longitudinal_gauge}
  ds^2 &= -(1+2\Phi) dt^2 + a(t)^2 (1-2\Phi)\delta_{ij}dx^i dx^j, & \phi(t,\mathbf{x}) &= \phi(t)+\delta\phi(t,\mathbf{x}), & \chi(t,\mathbf{x}) &= \chi(t)+\delta\chi(t,\mathbf{x}).
\end{align}
The linearised Klein--Gordon equations for $\delta\phi$ and $\delta\chi$ are coupled both to each other and to the metric perturbation $\Phi$:
\begin{align}\label{eq:section5_delta_field_equations}
  \begin{split}
    \delta\ddot\phi &+ 3H\delta\dot\phi + \left[\cfrac{k^2}{a^2}+\mathcal{V}_{\phi\phi}-2b_\phi^2e^{2b}\dot\chi^2\right]\delta\phi + \mathcal{V}_{\phi\chi}\delta\chi -2b_\phi e^{2b}\dot\chi\delta\dot\chi = 4\dot\phi\dot\Phi -2\mathcal{V}_\phi\Phi, \\
    \delta\ddot\chi &+ \left(3H+2b_\phi\dot\phi\right)\delta\dot\chi + \left(\cfrac{k^2}{a^2}+e^{-2b}\mathcal{V}_{\chi\chi}\right)\delta\chi +2b_\phi\dot\chi\delta\dot\phi + e^{-2b}\left(\mathcal{V}_{\chi\phi}-2b_\phi\mathcal{V}_\chi\right)\delta\phi = 4\dot\chi\dot\Phi -2e^{-2b}\mathcal{V}_\chi\Phi .
  \end{split}
\end{align}
The metric perturbation is not an independent scalar degree of freedom. The linearised metric equations give the constraints
\begin{align}\label{eq:section5_metric_constraints}
  \begin{split}
    3H(\dot\Phi+H\Phi)+\dot H\Phi+\cfrac{k^2}{a^2}\Phi
    &=-\cfrac{1}{2\mP^2}\left[\dot\phi\delta\dot\phi+e^{2b}\dot\chi\delta\dot\chi+b_\phi e^{2b}\dot\chi^2\delta\phi+\mathcal{V}_\phi\delta\phi+\mathcal{V}_\chi\delta\chi\right], \\
    \dot\Phi+H\Phi
    &=\cfrac{1}{2\mP^2}\left[\dot\phi\delta\phi+e^{2b}\dot\chi\delta\chi\right].
  \end{split}
\end{align}
The field system \eqref{eq:section5_delta_field_equations} and the constraints \eqref{eq:section5_metric_constraints} show explicitly how the two field perturbations and the scalar metric perturbation are coupled before gauge-invariant variables are introduced.

The field perturbations $\delta\phi$ and $\delta\chi$, as well as $\Phi$, are gauge dependent. We therefore introduce the gauge-invariant Mukhanov--Sasaki variables
\begin{align}\label{eq:section5_MS_variables}
  Q_\phi &= \delta\phi + \cfrac{\dot\phi}{H} \, \Phi, &
  Q_\chi &= \delta\chi + \cfrac{\dot\chi}{H} \, \Phi .
\end{align}
These variables coincide with the field perturbations in the spatially flat gauge. In particular, they show that the perturbation problem is already genuinely two-field before any adiabatic-entropy rotation is performed. The corresponding field equations read
\begin{align}\label{eq:section5_MS_equations}
  \begin{split}
    \ddot{Q_\phi} + 3H\dot{Q_\phi} -2e^{2b}b_\phi\dot\chi\dot{Q_\chi} + \left(\cfrac{k^2}{a^2}+C_{\phi\phi}\right)Q_\phi + C_{\phi\chi}Q_\chi &= 0, \\
    \ddot{Q_\chi} + \left(3H+2b_\phi\dot\phi\right)\dot{Q_\chi} +2b_\phi\dot\chi\dot{Q_\phi} + \left(\cfrac{k^2}{a^2}+C_{\chi\chi}\right)Q_\chi + C_{\chi\phi}Q_\phi &= 0,
  \end{split}
\end{align}
where the background-dependent coefficients are
\begin{align}\label{eq:section5_MS_coefficients}
  \begin{split}
    C_{\phi\phi} =& -2e^{2b}b_\phi^2\dot\chi^2 + \cfrac{3\dot\phi^2}{\mP^2} - \cfrac{e^{2b}\dot\phi^2\dot\chi^2}{2\mP^4H^2} - \cfrac{\dot\phi^4}{2\mP^4H^2} + \cfrac{2\dot\phi\mathcal{V}_\phi}{\mP^2H} + \mathcal{V}_{\phi\phi}, \\
    C_{\phi\chi} =& \cfrac{3e^{2b}\dot\phi\dot\chi}{\mP^2} - \cfrac{e^{4b}\dot\phi\dot\chi^3}{2\mP^4H^2} - \cfrac{e^{2b}\dot\phi^3\dot\chi}{2\mP^4H^2} + \cfrac{\dot\phi\mathcal{V}_\chi}{\mP^2H} + \cfrac{e^{2b}\dot\chi\mathcal{V}_\phi}{\mP^2H} + \mathcal{V}_{\phi\chi}, \\
    C_{\chi\chi} =& \cfrac{3e^{2b}\dot\chi^2}{\mP^2} - \cfrac{e^{4b}\dot\chi^4}{2\mP^4H^2} - \cfrac{e^{2b}\dot\phi^2\dot\chi^2}{2\mP^4H^2} + \cfrac{2\dot\chi\mathcal{V}_\chi}{\mP^2H} + e^{-2b}\mathcal{V}_{\chi\chi}, \\
    C_{\chi\phi} =& \cfrac{3\dot\phi\dot\chi}{\mP^2} - \cfrac{e^{2b}\dot\phi\dot\chi^3}{2\mP^4H^2} - \cfrac{\dot\phi^3\dot\chi}{2\mP^4H^2} -2e^{-2b}b_\phi\mathcal{V}_\chi + \cfrac{e^{-2b}\dot\phi\mathcal{V}_\chi}{\mP^2H} + \cfrac{\dot\chi\mathcal{V}_\phi}{\mP^2H} + e^{-2b}\mathcal{V}_{\phi\chi} .
  \end{split}
\end{align}
Thus, already at the level of the Mukhanov--Sasaki equations, the two scalar modes remain non-trivially coupled through both the kinetic prefactor and the effective potential.

For the scalar analysis, it is more useful to decompose the Mukhanov--Sasaki variables into the instantaneous adiabatic and entropy directions in field space. Following~\cite{Lalak:2007vi}, we define the instantaneous adiabatic mode $Q_\sigma$ and the instantaneous entropy mode $\delta s$:
\begin{align}\label{eq:section5_adiabatic_entropy_variables}
  Q_\sigma &= \cos\theta\, Q_\phi + \sin\theta \, e^b Q_\chi, & \delta s &= -\sin\theta \, Q_\phi + \cos\theta \, e^b Q_\chi,
\end{align}
with
\begin{align}\label{eq:section5_theta_definitions}
  \cos\theta &= \cfrac{\dot\phi}{\dot\sigma}, & \sin\theta &= \cfrac{e^b\dot\chi}{\dot\sigma}, & \dot\sigma &= \sqrt{\dot\phi^2+e^{2b}\dot\chi^2} .
\end{align}
The perturbation equations take the form
\begin{align}\label{eq:perturbation_equations}
  \begin{split}
    \ddot{Q_\sigma} + 3H\dot{Q_\sigma} + \left(\cfrac{k^2}{a^2}+C_{\sigma\sigma}\right)Q_\sigma + 2 \, \cfrac{\mathcal{V}_s}{\dot\sigma} \, \dot{\delta s}+C_{\sigma s}\delta s &=0, \\
    \ddot{\delta s}+3H\dot{\delta s}+\left(\cfrac{k^2}{a^2}+C_{ss}\right)\delta s -2 \, \cfrac{\mathcal{V}_s}{\dot\sigma} \, \dot{Q_\sigma}+C_{s\sigma}Q_\sigma &=0 .
  \end{split}
\end{align}
The terms proportional to $\mathcal{V}_s/\dot\sigma$ encode the transfer between adiabatic and entropy modes. This is the system that we solve numerically below.

The background-dependent coefficients are
\begin{align}\label{eq:section5_adiabatic_entropy_coefficients}
  \begin{split}
    C_{\sigma\sigma} &= \mathcal{V}_{\sigma\sigma} - \left( \cfrac{\mathcal{V}_{s}}{\dot\sigma} \right)^2 + \cfrac{2\dot\sigma\mathcal{V}_{\sigma}}{\mP^2H} + \cfrac{3\dot\sigma^2}{\mP^2} - \cfrac{\dot\sigma^4}{2\mP^4H^2} - b_{\phi} \left( s_\theta^2c_\theta\mathcal{V}_{\sigma} + (c_\theta^2+1)s_\theta\mathcal{V}_{s} \right), \\
    C_{\sigma s} &= 6H\cfrac{\mathcal{V}_{s}}{\dot\sigma} + \cfrac{2\mathcal{V}_{\sigma}\mathcal{V}_{s}}{\dot\sigma^2} +2\mathcal{V}_{\sigma s}+\cfrac{\dot\sigma\mathcal{V}_{s}}{\mP^2H}+2b_\phi\left(s_\theta^3\mathcal{V}_{\sigma}-c_\theta^3\mathcal{V}_{s}\right), \\
    C_{ss} &= \mathcal{V}_{ss} - \left( \cfrac{\mathcal{V}_{s}}{\dot\sigma} \right)^2 + b_\phi(1+s_\theta^2)c_\theta\mathcal{V}_{\sigma}+b_\phi c_\theta^2s_\theta\mathcal{V}_{s} - \dot\sigma^2\left(b_{\phi\phi}+b_\phi^2\right), \\
    C_{s\sigma} &= -6H\cfrac{\mathcal{V}_{s}}{\dot\sigma} - \cfrac{2\mathcal{V}_{\sigma}\mathcal{V}_{s}}{\dot\sigma^2} + \cfrac{\dot\sigma\mathcal{V}_{s}}{\mP^2H} .
  \end{split}
\end{align}
Here $s_\theta=\sin\theta$ and $c_\theta=\cos\theta$, while the projected derivatives of the potential are defined by
\begin{align}\label{eq:section5_projected_potential_derivatives}
  \begin{split}
    \mathcal{V}_{\sigma} &= \cos\theta\,\mathcal{V}_{\phi}+\sin\theta\,e^{-b}\mathcal{V}_{\chi}, \\
    \mathcal{V}_{s} &= -\sin\theta\,\mathcal{V}_{\phi}+\cos\theta\,e^{-b}\mathcal{V}_{\chi}, \\
    \mathcal{V}_{\sigma\sigma} &= \cos^2\theta\,\mathcal{V}_{\phi\phi}+2e^{-b}\sin\theta\cos\theta\,\mathcal{V}_{\phi\chi}+e^{-2b}\sin^2\theta\,\mathcal{V}_{\chi\chi}, \\
    \mathcal{V}_{\sigma s} &= -\sin\theta\cos\theta\,\mathcal{V}_{\phi\phi}+e^{-b}(\cos^2\theta-\sin^2\theta)\mathcal{V}_{\phi\chi}+e^{-2b}\sin\theta\cos\theta\,\mathcal{V}_{\chi\chi}, \\
    \mathcal{V}_{ss} &= \sin^2\theta\,\mathcal{V}_{\phi\phi}-2e^{-b}\sin\theta\cos\theta\,\mathcal{V}_{\phi\chi}+e^{-2b}\cos^2\theta\,\mathcal{V}_{\chi\chi} .
  \end{split}
\end{align}
Lastly, we define the gauge-invariant comoving curvature perturbation and the normalised entropy perturbation:
\begin{align}\label{eq:section5_R_S_definition}
  \mathcal{R} &= \cfrac{H}{\dot\sigma}\,Q_\sigma, &
  \mathcal{S} &= \cfrac{H}{\dot\sigma}\,\delta s .
\end{align}
These are the variables from which we extract the scalar power spectrum and determine the spectral tilt $n_s$.

The turn rate measures the bending of the homogeneous trajectory and controls the derivative coupling between the adiabatic and entropy perturbations. The coefficient $C_{ss}$ governs the local entropy-sector response in \eqref{eq:perturbation_equations}. A large turn rate or a negative $C_{ss}$ can therefore identify a transient interval in which the entropy mode affects the curvature mode, although neither quantity alone measures the accumulated transfer \cite{Gordon:2000hv,Lalak:2007vi}. For the numerical diagnostics, we define
\begin{align}\label{eq:section5_entropy_diagnostic_definitions}
  \Omega_{\rm turn} &\overset{\text{def}}{=} -\cfrac{\mathcal{V}_s}{\dot\sigma}, & m_{s,\rm eff}^2 &\overset{\text{def}}{=} C_{ss} .
\end{align}
Their extrema are taken over the complete integration interval from $N_0$ to the end of inflation. Therefore, they are diagnostics of potentially important transient dynamics, not direct measures of the integrated entropy-to-curvature transfer.

For each Fourier mode $k$, we solve the coupled system \eqref{eq:perturbation_equations} twice. In the first run, the canonically normalised adiabatic perturbation is placed in the Bunch--Davies state while the independent entropy perturbation is set to zero. In the second run, the roles are exchanged. After transforming these two independent solutions to $(\mathcal R,\mathcal S)$, we denote them by $(\mathcal R_1,\mathcal S_1)$ and $(\mathcal R_2,\mathcal S_2)$. Since the two initial quantum variables are statistically independent, the final spectra are obtained by summing the two contributions:
\begin{align}\label{eq:section5_power_spectra_definitions}
  \mathcal P_{\mathcal R}(k) &= \cfrac{k^3}{2\pi^2} \left(|\mathcal R_1|^2+|\mathcal R_2|^2\right), & \mathcal P_{\mathcal S}(k) &= \cfrac{k^3}{2\pi^2} \left(|\mathcal S_1|^2+|\mathcal S_2|^2\right).
\end{align}
To quantify the part of the final curvature power generated by the initially entropic solution, we also define
\begin{align}\label{eq:section5_entropy_seeded_curvature_fraction}
  \begin{split}
    \mathcal P_{\mathcal R}^{(\sigma)}(k) &= \cfrac{k^3}{2\pi^2} \, |\mathcal R_1|^2\, , \\
    \mathcal P_{\mathcal R}^{(s)}(k) &= \cfrac{k^3}{2\pi^2} \, |\mathcal R_2|^2\, , \\
    f_{s\to\mathcal R}(k) &\overset{\text{def}}{=} \cfrac{\mathcal P_{\mathcal R}^{(s)}(k)}{\mathcal P_{\mathcal R}^{(\sigma)}(k)+\mathcal P_{\mathcal R}^{(s)}(k)}\, .
  \end{split}
\end{align}
A second control evolution keeps the same background but removes the off-diagonal adiabatic-entropy couplings in~\eqref{eq:perturbation_equations}. It is used only as a numerical diagnostic, not as a separate physical model. We define
\begin{align}\label{eq:section5_no_mixing_diagnostic}
  \Delta_{\rm mix}(k) \overset{\text{def}}{=} \cfrac{\mathcal P_{\mathcal R}^{\rm full}(k)-\mathcal P_{\mathcal R}^{\rm no\,mix}(k)}{\mathcal P_{\mathcal R}^{\rm full}(k)}\, .
\end{align}
The two quantities answer different questions. The fraction $f_{s\to\mathcal R}$ decomposes the final curvature power according to the initially adiabatic and initially entropic solutions; by itself, it is not a direct measure of the dynamical effect of their coupling. The mixing-off comparison $\Delta_{\rm mix}$ directly measures how the off-diagonal adiabatic--entropy coupling changes the total curvature spectrum. They supplement the final entropy fraction and the extrema of the turn rate and entropy mass.
This procedure follows the two-field calculation of~\cite{Lalak:2007vi}, in which curvature and isocurvature modes are evolved numerically from well inside the Hubble radius to the end of inflation. Around the pivot scale, we fit the curvature spectrum by
\begin{align}\label{eq:section5_power_law_fit}
  \mathcal P_{\mathcal R}(k) = A_{\mathcal R} \left(\cfrac{k}{k_*}\right)^{n_s-1}.
\end{align}

We sample the interval around $k_*$ with $n_k$ Fourier modes. These modes are used both to integrate the coupled perturbation system and to fit the local power law \eqref{eq:section5_power_law_fit}. For the production calculation, we set
\begin{align}\label{eq:section5_numerical_benchmarks}
  m_0 &= 10^{-5}\mP, & \beta &= 1, & n_k &= 501 .
\end{align}
The value of $m_0$ is a benchmark of the correct inflationary order of magnitude rather than an exact CMB-normalised scalaron mass. A change of $m_0$ mainly rescales the absolute amplitude $A_{\mathcal R}$, whereas the comparison between the two-field runs and the Starobinsky reference is encoded in the relative amplitude shift and in the change of $n_s$. The entry $\beta=1$ specifies the benchmark coefficient of the complete theory and the tensor estimate. The reduced scalar mode equations \eqref{eq:perturbation_equations} do not depend directly on $\beta$ because the derivative interaction is omitted in the reduced calculation.

The background relaxation interval and the perturbation mode selection are treated separately. The common initial surface is the one defined in Section~\ref{section_simple_inflation}, while the pivot-like mode on each trajectory is defined by
\begin{align}\label{eq:section5_kstar_definition}
  k_* =& a(N_1)H(N_1)\, , & N_\text{rem}(N_1) &=60\, .
\end{align}
The spectral fit uses
\begin{align}\label{eq:section5_k_window}
  k\in[0.8k_*,1.2k_*] .
\end{align}
The numerical scan checks the initial ratio $C=k/(aH)$ for every fitted mode. Over all $246$ runs and all fitted modes, the shallowest production mode has
\begin{align}\label{eq:section5_minimum_BD_depth}
  C_\text{min}^{\rm prod}=2.3815\times10^3\, .
\end{align}
Thus the Bunch--Davies state is imposed deeply inside the Hubble radius. In addition to this admissibility guard, we verify convergence by restarting the exact-branch calculation at shallower surfaces. Table~\ref{tab:section5_BD_convergence} compares these runs with the production result. The distinction between specifying sub-Hubble vacuum data and checking independence of the chosen initial surface is standard in numerical multifield calculations \cite{Price:2014xpa}.

\begin{table}[t]
  \centering
  \renewcommand{\arraystretch}{1.15}
  \setlength{\tabcolsep}{7pt}
  \begin{tabular}{cccc}
    \hline
    $C_\text{in}$ & $\Delta A_{\mathcal R}/A_{\mathcal R}^{\rm St}$ & $\Delta n_s$ & $\max_k|\Delta\mathcal P_{\mathcal R}/\mathcal P_{\mathcal R}|$ \\
    \hline
    $50$   & $-4.77\times10^{-6}$ & $1.45\times10^{-4}$ & $3.81\times10^{-4}$ \\
    $100$  & $ 1.47\times10^{-8}$ & $1.43\times10^{-5}$ & $9.80\times10^{-5}$ \\
    $500$  & $ 5.72\times10^{-8}$ & $1.86\times10^{-7}$ & $3.97\times10^{-6}$ \\
    $1000$ & $ 4.57\times10^{-8}$ & $5.21\times10^{-8}$ & $1.12\times10^{-6}$ \\
    \hline
  \end{tabular}
  \caption{Initial-surface convergence of the exact Starobinsky model relative to the production surface, whose shallowest fitted mode has $C_\text{min}^{\rm prod}=2.3815\times10^3$.}
  \label{tab:section5_BD_convergence}
\end{table}
The $C_\text{in}=1000$ run changes the fitted amplitude and tilt by less than $5.3\times10^{-8}$ and the complete fitted spectrum by less than $1.2\times10^{-6}$. By contrast, $C_\text{in}=50$, although safely sub-Hubble, is not sufficient for the final quoted precision.

We scan the same transverse domain as in the background stability analysis,
\begin{align}\label{eq:section5_z_P_z_scan_domain}
  z(N_0),P_z(N_0)\in[-0.1,0.1],
\end{align}
using a uniform $7\times7$ grid. We also scan the mass ratio
\begin{align}\label{eq:section5_mu_scan_values}
  \mu\overset{\text{def}}{=}\cfrac{m_\chi}{m_0} \in \{0.90,0.95,1.00,1.05,1.10\}.
\end{align}
For $\mu\geq1$, the spectra scan and the uniform finite-amplitude contraction result apply to the same domain. The cases $\mu=0.90$ and $0.95$ test the lighter masses for which a small number of finite-amplitude initial directions do not contract over the reference interval. The production calculation contains $245$ two-field backgrounds and one pure Starobinsky reference. All $246$ runs completed successfully, and all $n_k=501$ Fourier modes were integrated successfully in every run. The perturbation solver, production configuration, logs, diagnostics, and output tables are associated with the Mendeley Data record \cite{Latosh2026MendeleyData}.

The pure Starobinsky calculation gives
\begin{align}\label{eq:section5_Starobinsky_reference_values}
  A_{\mathcal R}^{\rm St} &= 1.6449558488\times10^{-9}\, , & n_s^{\rm St} &= 0.9677549970\, .
\end{align}
The two-field runs with $z(N_0)=P_z(N_0)=0$ reproduce these curvature observables exactly for every scanned $\mu$, providing an internal check of the invariant branch. Table~\ref{tab:section5_envelope_mu_scan} gives the envelope over the $7\times7$ initial-data grid for each mass ratio.

\begin{table}[t]
  \centering
  \renewcommand{\arraystretch}{1.15}
  \setlength{\tabcolsep}{6pt}
  \begin{tabular}{ccccc}
    \hline
    $\mu$ & $\max|\Delta A_{\mathcal R}|/A_{\mathcal R}^{\rm St}$ & $\max|\Delta n_s|$ & $\max A_{\mathcal S}/A_{\mathcal R}$ & $\max f_{s\to\mathcal R}$ \\
    \hline
    $0.90$ & $1.46\times10^{-7}$ & $2.00\times10^{-7}$ & $4.51\times10^{-6}$ & $0.266703$ \\
    $0.95$ & $1.20\times10^{-7}$ & $1.87\times10^{-7}$ & $1.70\times10^{-6}$ & $0.266656$ \\
    $1.00$ & $1.22\times10^{-7}$ & $1.80\times10^{-7}$ & $5.84\times10^{-7}$ & $0.266607$ \\
    $1.05$ & $2.44\times10^{-7}$ & $1.78\times10^{-7}$ & $1.81\times10^{-7}$ & $0.266557$ \\
    $1.10$ & $3.67\times10^{-7}$ & $1.83\times10^{-7}$ & $4.93\times10^{-8}$ & $0.266506$ \\
    \hline
  \end{tabular}
  \caption{Envelope of the fully evolved reduced scalar spectra over the $7\times7$ transverse grid. The shifts are measured relative to \eqref{eq:section5_Starobinsky_reference_values}. The last column gives the largest entropy-seeded fraction over the fitted $k$-interval.}
  \label{tab:section5_envelope_mu_scan}
\end{table}

Over the entire scan,
\begin{align}\label{eq:section5_global_curvature_envelope}
  \max\cfrac{|\Delta A_{\mathcal R}|}{A_{\mathcal R}^{\rm St}} &=3.67\times10^{-7}\, , &
  \max|\Delta n_s| &=2.00\times10^{-7}\, , &
  \max\cfrac{A_{\mathcal S}}{A_{\mathcal R}} &=4.51\times10^{-6}\, .
\end{align}
The largest amplitude shift occurs for $\mu=1.10$ and $z(N_0)=P_z(N_0)=-0.1$, while the largest tilt shift occurs for $\mu=0.90$ at the same corner. These shifts are well below the numerical and observational precision relevant here.

The background coefficients nevertheless display genuine transient multifield dynamics:
\begin{align}\label{eq:section5_global_multifield_diagnostics}
  \max\cfrac{|\Omega_{\rm turn}|}{H} &=1.33464\, , &
  \min\cfrac{m_{s,\rm eff}^2}{H^2} &=-1.76985\, .
\end{align}
Both extrema occur for the displaced $\mu=1.10$ corner trajectories. They show that the entropy sector is not dynamically absent, but they do not measure accumulated transfer by themselves.

\begin{figure}[t]
  \centering
  \includegraphics[width=0.48\textwidth]{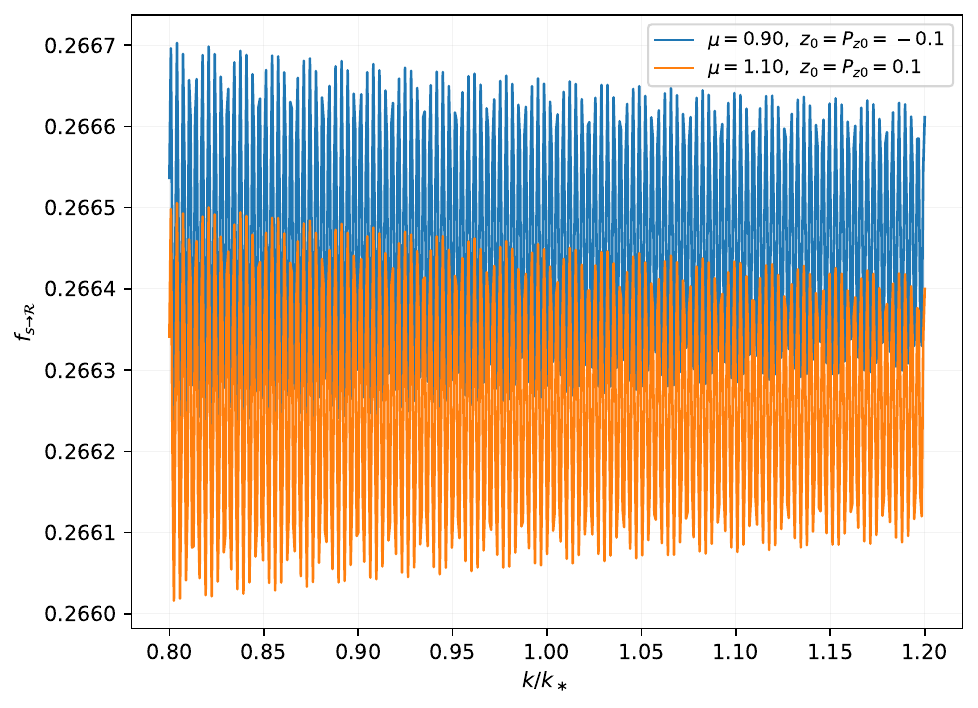}\hfill
  \includegraphics[width=0.48\textwidth]{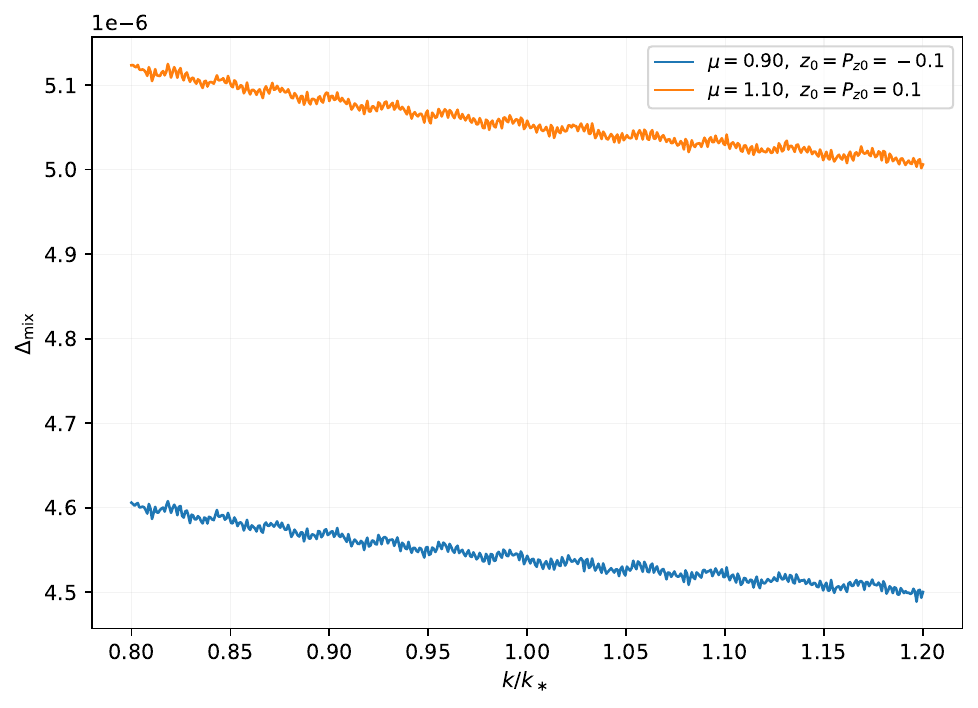}
  \caption{Complementary entropy diagnostics for representative equal-sign corner trajectories. Left: decomposition of the final curvature power according to the initially entropic solution. Right: direct diagnostic of the relative change in the total curvature spectrum when the off-diagonal adiabatic--entropy couplings are removed. The no-mixing evolution is a diagnostic modification of the perturbation equations, not a separate physical model.}
  \label{fig:section5_entropy_transfer}
\end{figure}

The two complementary entropy diagnostics resolve this point. The largest entropy-seeded fraction over the entire spectra scan is
\begin{align}\label{eq:section5_global_entropy_transfer}
  \max f_{s\to\mathcal R}=0.266703\, ,
\end{align}
attained for $\mu=0.90$ and $z(N_0)=P_z(N_0)=\pm0.1$. Thus, approximately $26.7\%$ of the final curvature power in the most displaced light-field trajectory is carried by the solution initially placed in the entropy direction. However, the no-mixing control gives
\begin{align}\label{eq:section5_global_no_mixing_result}
  \max|\Delta_{\rm mix}|=5.125\times10^{-6}\, ,
\end{align}
with the largest value at the $\mu=1.10$ equal-sign corners. The two quantities have complementary meanings: $f_{s\to\mathcal R}$ decomposes the final curvature power according to the initial adiabatic and entropy seeds, whereas $\Delta_{\rm mix}$ measures the change in the total spectrum when their coupling is removed. Therefore, the coupled evolution substantially redistributes the two independent seed contributions while leaving their total curvature power almost unchanged. A small final entropy spectrum alone would not establish this result. The direct mixing-off comparison does. Figure~\ref{fig:section5_entropy_transfer} displays both diagnostics for representative equal-sign corner trajectories.

Combining the reduced spectra with the complete matrix scan gives a branch-restricted but internally consistent result. For $\mu\geq1$, the same $7\times7$ sample exhibits uniform finite-amplitude contraction, a healthy complete principal perturbation operator for $|\beta|\leq1$, and an essentially Starobinsky total curvature spectrum. The lighter masses provide additional stress tests: their sampled square is not uniformly contracting over eight e-folds, but their perturbation spectra remain equally robust.

\section{Discussion and conclusion}\label{section_conclusion}

We analysed inflation in the minimal two-field scalar-tensor completion defined by the one-loop effective action~\eqref{the_effective_action}. The model contains the exact invariant Starobinsky surface $z=P_z=0$. Its homogeneous stability was tested using four-dimensional trajectories of the reduced slow-roll background system, the transverse variational equations, and the largest finite-time transverse Lyapunov exponent. On the $7\times7$ initial-data grid and over the reference eight-e-fold interval, the entire sampled square contracts uniformly for $\mu\geq1$. For $\mu=0.90$ and $0.95$, the linear transverse map still contracts, but four and two finite-amplitude grid points, respectively, do not satisfy $R_\perp<1$. These statements are local and finite-time.

The auxiliary-field representation of the derivative-coupling theory belongs to multifield generalised $G$-inflation and has second-order equations of motion. We derived the exact constraint-reduced tensor and scalar quadratic actions, including all lapse- and shift-induced terms, and evaluated them on all $246$ backgrounds for $\beta\in\{-1,-0.1,0,0.1,1\}$. The smallest kinetic eigenvalue is $0.9999824$, both scalar speeds remain within $4.0\times10^{-10}$ of unity, and the off-diagonal kinetic and gradient correlations remain below $4.2\times10^{-3}$. The kinetically normalised derivative mixing is below $8.5\times10^{-3}H$, and the tensor speed differs from unity by less than $9.8\times10^{-13}$. A transient kinetically normalised mass eigenvalue reaches approximately $-2.25H^2$, but it is not accompanied by a ghost or high-frequency gradient instability. The principal results are stable under matrix-grid refinement. Thus, perturbative stability is tested with the complete constraint-reduced quadratic operator over the stated $\beta$ range, whereas the production curvature spectra quoted below are obtained from the reduced scalar evolution.

The complete-spectrum comparison in $\beta$, together with the separate $\beta=0$ complete--reduced regression, gives the conservative bound
\[
  \max|\Delta_\beta|+\epsilon_{\rm numerical}
  +\epsilon_{\rm CPU\text{-}GPU}
  \leq \csname VALIDATION_COMPLETE_BOUND\endcsname,
\]
where the CPU solver supplies the reference production estimate and numerical-convergence allowance, while the dense GPU calculation supplies an independent full-domain cross-check. Reference status does not exclude the GPU result: the bound uses its larger observed correction and includes the full CPU--GPU discrepancy as an allowance. The independent common-basis calculation bounds $\|\Delta M_{\rm phys}\|_2/H^2$ and $\|\Delta\Omega_{\rm phys}\|_2/H$ by $\csname VALIDATION_DELTA_M\endcsname$ and $\csname VALIDATION_DELTA_OMEGA\endcsname$. These quantitative checks validate the omission of the derivative-coupling correction from the production scalar evolution at better than the $10^{-3}$ accuracy relevant to the physical conclusion, within the stated fixed reduced-background scope.

The reduced scalar perturbations were evolved as a genuinely coupled adiabatic--entropy system from a production surface with $C_\text{min}=2.3815\times10^3$ to the end of inflation. The Starobinsky reference is
\begin{align}
  A_{\mathcal R}^{\rm St} &=1.6449558488\times10^{-9}\, , & n_s^{\rm St} &=0.9677549970\, .
\end{align}
Across the five masses and the entire $7\times7$ transverse grid, the largest relative amplitude shift is $3.67\times10^{-7}$, and the largest tilt shift is $2.00\times10^{-7}$. Moving the initial surface to $C_\text{in}=1000$ changes the fitted amplitude and tilt by less than $5.3\times10^{-8}$ and the fitted spectrum by less than $1.2\times10^{-6}$.

The multifield mode composition is nontrivial despite the robustness of the total spectrum. The entropy-seeded solution contributes up to $26.7\%$ of the final curvature power, whereas switching off the adiabatic--entropy mixing changes the total curvature spectrum by at most $5.125\times10^{-6}$. The coupled evolution therefore redistributes power between the independent initial solutions without producing an observable change in their sum. This distinction is essential: the small final entropy spectrum does not imply the absence of intermediate entropy-to-curvature conversion.

Our conclusions apply to the stated five mass ratios, the $7\times7$ initial-data sample, the tested coefficient interval $|\beta|\leq1$, and the Starobinsky-like slow-roll branch. Other regions of phase space, kinetically driven solutions, larger derivative couplings, or microscopic interactions that generate additional scalar-tensor operators may behave differently.

\section*{Data and code availability}

The numerical project \texttt{starobinsky\_modification\_numerics} and the accompanying data package contain the resolved JSON configurations, background and stability runs, complete matrix diagnostics, reduced spectra, Bunch--Davies convergence tests, entropy-transfer calculations, logs, HDF5 result files, and the scripts used to regenerate the tables and figures. The published Mendeley Data record \cite{Latosh2026MendeleyData} contains the data associated with the original analysis; the final revision package supersedes the corresponding numerical tables with the unified calculation reported here.

\section*{Acknowledgements}
The work was supported by the Foundation for the Advancement of Theoretical Physics and Mathematics ``BASIS''. The author is grateful to \href{https://orcid.org/0000-0002-1073-3967}{Pavel Petrov} for fruitful discussions.

\bibliographystyle{unsrturl}
\bibliography{tSEoSI}

\end{document}